\documentclass[a4paper,fleqn,usenatbib]{mnras}

\usepackage{newtxtext,newtxmath}

\usepackage[T1]{fontenc}
\usepackage{ae,aecompl}

\usepackage{adjustbox}
\usepackage{graphicx}	
\usepackage{amsmath}	
\usepackage{amssymb}	
\usepackage{xcolor} 
\usepackage{multirow,booktabs,tabularx} 
\usepackage[font=small,labelfont=bf,tableposition=top]{caption}
\usepackage{threeparttable}

\graphicspath{{}}
\defcitealias{Yuan:2018ec}{Paper~I}
\defcitealias{Yoon2018}{Paper~II}

\title[Parameter Surveys of AGN Feedback]{Active Galactic Nucleus Feedback in an Elliptical Galaxy with the Most Updated AGN Physics: Parameter Explorations}

\author[Yao et al.]{
Zhiyuan Yao $^{1, 2}$\thanks{E-mail: yzy1116@shao.ac.cn},
Feng Yuan$^{1,2}$\thanks{E-mail: fyuan@shao.ac.cn}
and Jeremiah P. Ostriker$^{3,4}$
\\
$^{1}$Shanghai Astronomical Observatory, Chinese Academy of Sciences, 80 Nandan Road, Shanghai 200030, China\\
$^{2}$University of Chinese Academy of Sciences, 19A Yuquan Road, Beijing 100049, People's Republic of China\\
$^{3}$Department of Astronomy, Columbia University, 550 W. 120th Street, New York, NY 10027, USA\\
$^{4}$Department of Astrophysical Sciences, Princeton University, Princeton, NJ 08544, USA}

\date{Accepted XXX. Received YYY; in original form ZZZ}

\pubyear{2020}

\begin{document}
\label{firstpage}
\pagerange{\pageref{firstpage}--\pageref{lastpage}}
\maketitle

\begin{abstract}
In a previous work, we have proposed a sub-grid model of active galactic nucleus (AGN) feedback by taking into account the state-of-the-art AGN physics, and used that model to study the effect of AGN feedback on the evolution of an isolated elliptical galaxy by performing two dimensional high-resolution (i.e., the Bondi radius is well resolved) simulations. In that work, typical values of model parameters were adopted. In the present work, we extend that study by exploring the effects of uncertainties of parameter values. Such a study is also useful for us to understand the respective roles of various components of the model. These parameters include the mass flux and velocity of AGN wind and radiative efficiency in both the hot and cold feedback modes, and the initial black hole (BH) mass.  We find that the velocity of AGN wind in the hot mode is the most important quantity to control the typical accretion rate and luminosity of AGN, and the mass growth of the BH. 
The effect of the wind on star formation is less sensitive. Within the limited parameter range explored in the current work, a stronger AGN wind suppresses star formation within $\sim 100$~pc but enhances star formation beyond this radius, while the star formation integrated over the evolution time and the whole galaxy roughly remain unchanged. AGN radiation suppresses the BH accretion in a mild way, but dust is not considered here. Finally, a smaller initial BH mass results in a more violent evolution of the BH accretion rate. The corresponding AGN spends more time in the high-luminosity state and the percentage of BH mass growth is higher. Our results indicate the robustness of AGN feedback in keeping the galaxy quenched.
\end{abstract}




\section{Introduction}
\label{sec:introduction}
In the centre of every massive galaxy with a bulge there exists a supermassive BH (see, e.g.,\citealt{Kormendy_Coevolution_2013} for a review). Observations have found tight correlations between the mass of the BH and the properties of the classical bulge, including its stellar mass \citep{Magorrian1998,2004ApJ...604L..89H,Kormendy_Coevolution_2013}, luminosity \citep{2003ApJ...589L..21M,2009ApJ...698..198G}, and stellar velocity dispersion \citep{Gebhardt2000,2000ApJ...539L...9F,2002ApJ...574..740T}. These correlations suggest the coevolution of the central BH and its host galaxy, and AGN feedback might play a major role in this coevolution.

The literatures regarding AGN feedback has increased greatly in the past twenty years \citep{2018NatAs...2..198H}. Due to the complexities of this relatively young topic, AGN feedback is studied mainly by numerical simulations. The gap of the physical scales between the BH and the galaxy can be as large as nine orders of magnitude, indicating all the current simulations require sub-grid assumptions and approximations. \citet{DiMatteo2005} and \citet{Springel2005} first studied AGN feedback by using hydrodynamical cosmological simulations. Although the AGN feedback was simply employed through a free parameter (the feedback efficiency $\varepsilon_\mathrm{f}$, which determines the fraction of the AGN bolometric luminosity that couples to the gas near the BH), they found a tight correlation between the BH mass and the stellar velocity dispersion. On the other hand, by including AGN feedback, both semi-analytic modeling \citep{Croton2006,Bower2006,2007MNRAS.375.1189M} and hydrodynamical cosmological simulations \citep{2014MNRAS.444.1518V,2014MNRAS.444.1453D,2015MNRAS.450.1349K,2015MNRAS.446..521S,2018MNRAS.473.4077P} obtained much better fittings to observations, e.g., the galaxy luminosity function at the massive end of the galaxy mass, the ``downsizing problem'', and the ``cooling flow problem''. Therefore, it is generally believed that AGN feedback does play an important role in the galaxy evolution.

One of the most important quantities for the study of AGN feedback is the mass accretion rate of the BH, because it determines the power of AGN. However, in cosmological simulations, the scales relevant to the BH accretion and the ejection of AGN winds are not directly resolved due to resolution limitation. The BH accretion rate usually is calculated by the gas $\sim1$~kpc from the central BH, which could overestimate or underestimate the real BH accretion rate with the uncertainty as large as $\sim 300$ \citep{Negri2017}. Moreover, the AGN outputs such as radiation and matter outflows are often not speficied and the interactions between these AGN outputs and the gas in the galaxy are usually treated in a phenomenological  parameterized approach.

Since AGN feedback occurs in a single galaxy rather than cosmological scale, to investigate the details of how AGN feedback works, a perhaps better approach is zooming in on a  galaxy with high resolution which can resolve the Bondi radius $r_\mathrm{B}=2GM_\mathrm{BH}/c_\mathrm{s,\infty}^2$, which is roughly ten pc for the BH mass of $10^9~M_\odot$ and gas temperature of $10^8~$K (\citealt{Ciotti1997,Ciotti2001,Ciotti2007,Ciotti2009,Shin2010,Ciotti2010,Novak2011,Gan2014,2017ApJ...835...15C}; \citealt{Yuan:2018ec}, hereafter \citetalias{Yuan:2018ec}; \citealt{2019MNRAS.482.4642Z}). Within the Bondi radius, the gravity is dominated by the BH, so this is the regime of accretion flow.

Various types of accretion disks have been well investigated by the accretion disk community, including cold accretion mode at high accretion rate regime \citep[e.g.][]{2002apa..book.....F} and hot accretion mode in the low accretion rate regime \citep[e.g.][]{2014ARA&A..52..529Y}. The cold accretion mode correspond to radiative or quasar feedback mode, while the hot accretion mode corresponds to maintenance or radio or kinetic feedback mode. Some of these names are confusing or even misleading so in \citetalias{Yuan:2018ec} we suggest to call them cold and hot feedback modes respectively. Since the Bondi radius is commonly regarded as the outer boundary of accretion flow, once it is resolved as it is in our simulations, we can precisely calculate rather than estimate the accretion rate. Given the BH accretion rate, we can then adopt the accretion knowledge to calculate the outputs, including radiation and matter outflow, and further calculate their interaction with the gas in the host galaxy.

Taking an isolated elliptical galaxy as an example, \citetalias{Yuan:2018ec} investigated the effects of AGN feedback by establishing a sub-grid model of AGN feedback with state-of-the-art accretion physics. The inner radius of the simulation domain is several times smaller than the Bondi radius. The mass accretion rate is thus precisely calculated and the accretion (and feedback) mode can be determined.  Both radiation and momentum-driven wind were considered in the two modes. The properties of the wind in the cold mode were adopted from observations, while in the hot mode they were described based on the 3D GRMHD simulation results of \citet{2015ApJ...804..101Y} due to the rarity of observational data. The jet in the hot mode was temporarily omitted.

In \citetalias{Yuan:2018ec}, they focus on the case of low angular momentum of the galaxy. They examined the respective roles of radiation and wind feedback, and found that both can suppress star formation and cause the variability of the AGN, but the wind was by momentum interaction while radiation was by radiative heating. Wind was believed to play a more important role in suppressing star formation and the BH accretion rate. 
In the second paper of this series, \citet{Yoon2018} \citepalias{Yoon2018} extended this model to the high angular momentum case of the elliptical galaxy. They found that while some results were qualitatively similar to those in \citetalias{Yuan:2018ec}, other results, such as star formation and black hole growth, showed a significant difference due to the mass concentration in the galactic disk as a consequence of galactic rotation. More recently, \citet{2019ApJ...885...16Y} specifically focused on the role of hot mode feedback. They find that although the AGN power in the hot mode is much lower compared to the cold mode, the hot mode feedback still plays an important role,  because most time of the AGN stays in this mode.

One remaining question in \citetalias{Yuan:2018ec} is that, since the parameters of the sub-grid model  have some uncertainties, to what extent do these uncertainties effect the galaxy evolution. We will accomplish this goal in this paper.  The paper is structured as follows. In Section~\ref{sec:models}, we briefly introduce the framework of our models, including the galaxy setups, stellar feedback, star formation, and AGN feedback. In Section~\ref{sec:parameter} we introduce the parameters explored in this paper. Then in Section~\ref{sec:results}, we show our results. Finally, in Section \ref{sec:discussions}, we summarize our results and compare with the previous works.

\section{Models}
\label{sec:models}
In this section, we briefly introduce the main physical processes included in our models. The simulation begins with a massive elliptical galaxy at the age of 2~Gyr.  As in \citetalias{Yuan:2018ec} we adopt the sub-grid physics that divides the accretion and feedback into two modes depending on the accretion rate at the inner boundary. In each mode we consider both wind and radiation, which are then injected into the simulation region through inner radial boundary. We simulate the interactions of wind and radiation with the interstellar medium (ISM) by considering simplified radiative transfer. Stellar processes such as star formation, stellar winds, supernovae (SNe) Ia and II are taken into account as well. Readers are referred to \citetalias{Yuan:2018ec} for more information.

\subsection{Galaxy Model}
We focus on the secular evolution of a massive isolated elliptical galaxy. The gravitational potential we adopt is dominated by the dark matter halo beyond 10~kpc, by stars from 0.05 to 10~kpc, and by the central BH within 50~pc, respectively. 

Both the dark matter and stellar components are modeled through static potentials, and the potential from newly formed stars can be neglected since they are minor compared to other components. We adopt the Jaffe stellar distribution \citep{1983MNRAS.202..995J} embedded in the dark matter halo so that the total density satisfies the isothermal sphere assumption and decreases as $r^{-2}$ \citep{Ciotti2009}. The Jaffe stellar distribution is described by
\begin{equation}
\rho_\star = \frac{M_\star r_\star}{4\pi r^2(r_\star+r)^2},
\end{equation}
where $M_\star$ and $r_\star$ are the total stellar mass and the scale-length of the galaxy, respectively. $M_\star$ is set to be $3\times10^{11}~M_\odot$ and $r_\star$ is the scale length that corresponds to the effective radius $R_\mathrm{e}=0.7447~r_\star$. The total density profile is given by
\begin{equation}
\rho_\mathrm{T}=\frac{\sigma_0^2}{2\pi Gr^2},
\end{equation}
where $\sigma_0=260$~km s$^{-1}$ is the central projected velocity dispersion. From the Faber-Jackson relation and the Fundamental Plane, we can derive the total B-band luminosity $L_\mathrm{B}=5\times10^{10}~L_\mathrm{B,\odot}$ and the effective radius $R_\mathrm{e}=6.9$~kpc. The initial mass of the central BH is determined according to the empirical correlation $M_\mathrm{BH}/(10^9~M_\odot)=0.49~(M_\star/10^{11}~M_\odot)^{1.17}$ in \citet{Kormendy_Coevolution_2013}, thus yielding the BH mass $M_\mathrm{BH}=1.8\times10^9~M_\odot$ for $M_\star=3\times10^{11}~M_\odot$.

The initial gas fraction is negligible, and most of the gas is provided by stellar evolution in our work.
The galaxy is assumed to be slowly rotating, which is supported by observations that suggest slow rotators start appearing when $M_\star>2\times10^{11}~M_\odot$ \citep{Cappellari2013,2018MNRAS.477.4711G}. Since the stars are slowly rotating and the stellar wind is the only mass source in our simulations, the angular momentum of the gas is therefore low and we do not need to handle the angular momentum transfer (for the case of high angular momentum, readers are referred to \citetalias{Yoon2018}; \citet{Gan2019}). To focus on the effect of model parameters, the gaseous halo and cosmological inflow are not considered in this paper, consistent with \citetalias{Yuan:2018ec}.

\subsection{Star formation and Stellar Feedback}
\label{subsec:sf}
We implement star formation by subtracting mass, momentum, and energy from the grid. We note that stellar populations and their dynamics are not explicitly tracked in the simulation. Different from \citetalias{Yuan:2018ec}, the star formation is triggered only if the temperature is lower than $4\times10^{4}$~K and the number density is higher than $1$~cm$^{-3}$ concurrently. The aim is to mimic the surface density threshold of $\sim10~\mathrm{M}_\odot\mathrm{pc}^{-2}$ for star formation revealed by observations \citep{1989ApJ...344..685K,1998ApJ...498..541K,2001ApJ...555..301M,2008AJ....136.2846B}. The star formation rate per unit volume is given by the Kennicutt-Schmidt prescription
\begin{equation}
\dot{\rho}_\mathrm{SF}=\frac{\eta_\mathrm{SF}\rho}{\tau_\mathrm{SF}}.
\end{equation}
Here the star formation efficiency is $\eta_\mathrm{SF}=0.01$, which is an order of magnitude smaller than in  \citetalias{Yuan:2018ec}. This modification is driven by the observations on surface density relationships of galaxies \citep{1998ApJ...498..541K}, and local giant molecular clouds (\citealt{2007ApJ...654..304K}; for a comprehensive discussion see the review by \citealt{2019ARA&A..57..227K}). The star formation timescale is
\begin{equation}
\tau_\mathrm{SF}=\mathrm{max}(\tau_\mathrm{cool},\tau_\mathrm{dyn}).
\end{equation}
The cooling and the dynamical timescales are given by
\begin{equation}
\tau_\mathrm{cool}=\frac{E}{C}, \tau_\mathrm{dyn}=\mathrm{min}(\tau_\mathrm{Jeans}, \tau_\mathrm{rot})
\end{equation}
with
\begin{equation}
\tau_\mathrm{Jeans}\equiv\sqrt{\frac{3\pi}{16G\rho}}, \tau_\mathrm{rot}\equiv \frac{2\pi r}{v_\mathrm{K}(r)}.
\end{equation}
Here $v_\mathrm{K}(r)$, $E$, and $C$ are the Keplerian velocity at radius $r$, internal energy density, and net cooling rate per unit volume, respectively. We adopt the formulae in \citet{2005MNRAS.358..168S} to compute cooling, which includes the bremsstrahlung cooling, Compton cooling, line and recombination continuum cooling. 

We note that, as in many similar numerical simulation works of AGN feedback, our calculation of the star formation rate has big uncertainties. It is technically difficult to simulate the process of star formation from first principle in such large-scale simulations. These uncertainties and difficulties, including our neglect of the self-gravity of the ISM, can be absorbed in some degree in our simplified and parameterized treatment of our calculation of star formation rate.

The evolving stars inject mass and energy to the ISM, mainly during the asymptotic giant brach (AGB) phase. At the end of their lives, drastic outbursts called supernova feedback produce a large amount of energy and return most of their mass to the ISM. Based on the population of stars, SNe Ia and II are distinguished. All of these stellar feedback processes are considered in our simulations and the detailed descriptions can be found in \citet{2012ASSL..378...83C}. The chemical evolution and dust absorption are ignored for simplicity.

\subsection{AGN feedback}
\label{sec:agnfb}
The AGN feedback is divided into two modes depending on the accretion rate at the innermost grid radius. The boundary of the accretion rate can be inferred from the observations on the state transition of black hole X-ray binaries, which occurs under the critical luminosity $L_\mathrm{C}\sim2\%L_\mathrm{Edd}$, or the equivalent critical accretion rate $\dot{M}_\mathrm{C}\sim2\%\dot{M}_\mathrm{Edd}$. In principle, this critical accretion rate applies to the accretion rate at the BH horizon $\dot{M}_\mathrm{BH}$. Yet in practice, we simply rely on the accretion rate at the inner boundary $\dot{M}(r_\mathrm{in})$ to judge the feedback mode, which is calculated by
\begin{equation}
\dot{M}(r_\mathrm{in}) = 2\pi r_\mathrm{in}^2\int_0^\pi\rho(r_\mathrm{in},\theta)\mathrm{min}[v_\mathrm{r}(r_\mathrm{in},\theta),0]\mathrm{sin}\theta d\theta.
\end{equation}

Both radiation and wind are taken into account in each mode. The radiative transport is considered in a approximated way by assuming the flow is optically thin \citet{2012ASSL..378...83C}. The heating terms include the Compton heating and photoionization heating driven by the central AGN. The radiation pressure is included by considering both the electron scattering and the absorption of photons by atomic lines. Wind is input via momentum, which was found to be more powerful than the thermally driven wind and better consistent with observations by previous works \citep{2012ApJ...754..125C,2015MNRAS.449.4105C}.

In the hot mode, the radiative efficiency used when calculating the radiation flux of AGN is a function of accretion rate \citep{Xie:2012dv},
\begin{equation}
\varepsilon_\mathrm{hot}(\dot{M}_\mathrm{BH}) =\frac{\varepsilon_\mathrm{cold}}{0.057}\varepsilon_0\left(\frac{\dot{M}_\mathrm{BH}}{10^{-2}~\dot{M}_\mathrm{Edd}}\right)^a.
\end{equation}
Here $\varepsilon_\mathrm{cold}/0.057$ accounts for the spin of the BH and $\dot{M}_\mathrm{Edd}=L_\mathrm{Edd}/(\varepsilon_\mathrm{cold}c^2)$ is the Eddington accretion rate. The values of $\varepsilon_0$ and $a$ depend on $\dot{M}_\mathrm{BH}$ and $\delta$, which denotes the fraction of the viscously dissipated energy that directly heats electrons. Assuming $\delta=0.1$, $\varepsilon_0$ and $a$ are given by
\begin{equation}
\begin{aligned}
(\varepsilon_0, a) =
\left\{
	\begin{array}{ll}
         (0.12, 0.59), & \dot{M}_\mathrm{BH}/\dot{M}_\mathrm{Edd}\lesssim9.4\times10^{-5} \\
         (0.026, 0.27),& 9.4\times10^{-5}\lesssim\dot{M}_\mathrm{BH}/\dot{M}_\mathrm{Edd}\lesssim5\times10^{-3}\\
         (0.5, 4.53), & 5\times10^{-3}\lesssim\dot{M}_\mathrm{BH}/\dot{M}_\mathrm{Edd}\lesssim6.6\times10^{-3}\\
         (0.057, 0). &  6.6\times10^{-3}\lesssim\dot{M}_\mathrm{BH}/\dot{M}_\mathrm{Edd}\lesssim2\times10^{-2}\\
         \end{array}
\right.
\end{aligned}
\end{equation}
The Compton temperature in the hot mode, used to calculate the radiative heating by AGN to  the ISM, is calculated based on the spectral energy distribution of low-luminosity AGNs combined from the literature \citep{2017ApJ...844...42X} and is given by
\begin{equation}
\begin{aligned}
T_\mathrm{C,hot}=
\left\{
	\begin{array}{ll}
	10^8\mathrm{~K}, & 10^{-3}\lesssim \dot{M}_\mathrm{BH}/\dot{M}_\mathrm{Edd}\lesssim0.02 \\
	5\times10^7\mathrm{~K}. & \dot{M}_\mathrm{BH}/\dot{M}_\mathrm{Edd}\lesssim10^{-3}
	\end{array}
\right.
\end{aligned}
\end{equation}

In the hot mode, the accretion flow consists of an inner hot accretion flow plus an outer truncated thin disk. The truncation radius $r_\mathrm{tr}$ is described by (Yuan \& Narayan 2014)
\begin{equation}
r_\mathrm{tr}=3r_\mathrm{s}\left[\frac{2\times10^{-2}\dot{M}_\mathrm{Edd}}{\dot{M}(r_\mathrm{in})}\right]^2,
\end{equation}
where $r_\mathrm{s}$ is the Schwarzschild radius. The existence of a strong wind in hot accretion flows has been shown in \citet{2012ApJ...761..130Y}. Based on 3D GRMHD numerical simulation of black hole accretion, using the ``virtual particle trajectory" approach, \citet{2015ApJ...804..101Y} derived the mass flux and velocity of the wind:
\begin{equation}
\dot{M}_\mathrm{W,H}=\dot{M}_\mathrm{BH}\left(\frac{r_\mathrm{tr}}{20~r_\mathrm{s}}\right),
\end{equation}
\begin{equation}
v_\mathrm{W,H}=0.2v_\mathrm{K}(r_\mathrm{tr}),
\end{equation}
where $v_\mathrm{K}(r_\mathrm{tr})$ is the Keplerian velocity at $r_\mathrm{tr}$. The velocity is $\sim2000~$km/s when $\dot{M}(r_\mathrm{in})=10^{-3}~\dot{M}_\mathrm{Edd}$, and it increases to $\sim0.08~c$ as the $\dot{M}(r_\mathrm{in})$ approaches $0.02~\dot{M}_\mathrm{Edd}$. Given
\begin{equation}
\dot{M}(r_\mathrm{in}) = \dot{M}_\mathrm{BH}+\dot{M}_\mathrm{W,H},
\end{equation}
we can obtain the properties of the BH accretion rate and wind properties in the hot mode. Compared to jets, the opening angle of the wind is much larger, lying within $\theta\sim30\degr-70\degr$ and $110\degr-150\degr$ above and below the equatorial plane, respectively \citep{2015ApJ...804..101Y}. The mass flux of wind within the abovementioned two $\theta$-ranges is assumed to be independent of angles.

In the cold mode, the accretion rate is high and a standard thin disk model is applied. The radiative efficiency is commonly assumed $\varepsilon_\mathrm{cold}=0.1$ \citep{2002MNRAS.335..965Y,Marconi2004}, which means the BH is moderately spinning according to the thin disk model. The Compton temperature $T_\mathrm{C,cold}$, which measures the average energy of the emitting photons, is calculated from the observed spectrum of quasars \citep{2004MNRAS.347..144S} and is given by $2\times10^7$~K.

The wind properties in the cold mode are obtained from \citet{2015MNRAS.451.4169G}. They analyzed a sample of 51 \textit{Suzaku}-observed AGNs and independently detected Fe K absorption in 40 percent of the sample. After processing the data, they measured the mass flux and velocity of the wind:
\begin{equation}
\label{eq:coldwfx}
\dot{M}_\mathrm{W,C}=0.28\left(\frac{L_\mathrm{bol}}{10^{45}~\mathrm{erg~s}^{-1}}\right)^{0.85}~M_\odot~\mathrm{yr}^{-1},
\end{equation}
\begin{equation}
v_\mathrm{W,C}=2.5\times10^4\left(\frac{L_\mathrm{bol}}{10^{45}~\mathrm{erg~s}^{-1}}\right)^{0.4}~\mathrm{km~s}^{-1}.
\end{equation}
Obviously, the wind should not be isotropic, but the exact description of the distribution of the wind flux is still poorly constrained. Following the previous works \citep{Novak2011,Gan2014,2017ApJ...835...15C}, the mass flux of the wind is assumed to be proportional to cos$^2(\theta)$.

To obtain the properties of the wind in the cold mode, we need to calculate the BH accretion rate and  the AGN luminosity. Once the gas reaches the circularization radius $R_\mathrm{cir}$, the accretion disk is formed. With the total mass of the gas in the disk, $M_\mathrm{dg}$, the mass inflow rate at $R_\mathrm{cir}$ in the disk can be estimated as
\begin{equation}
\label{eq:dinflow}
\dot{M}_\mathrm{d,inflow}=\frac{M_\mathrm{dg}}{\tau_\mathrm{vis}},
\end{equation}
where the viscous timescale $\tau_\mathrm{vis}$ is described by \citep{2008bhad.book.....K}
\begin{equation}
\label{eq:vis}
\tau_\mathrm{vis}\approx1.2\times10^6~\mathrm{yr}\left(\frac{\alpha}{0.1}\right)^{-1}\left(\frac{R_\mathrm{cir}}{100~r_\mathrm{s}}\right)^{7/2}\left(\frac{M_\mathrm{BH}}{10^9M_\odot}\right).
\end{equation}
Here $\alpha$ is the viscosity parameter. Given
\begin{equation}
\dot{M}_\mathrm{d,inflow}=\dot{M}_\mathrm{BH}+\dot{M}_\mathrm{W,C},
\end{equation}
we can obtain the BH accretion rate and the wind properties in the cold mode. We would like to point out that, when $\dot{M}(r_{\rm in})$ is above $2\%\dot{M}_{\rm Edd}$, the above approach of calculating $\dot{M}_{\rm BH}$ does not ensure $\dot{M}_{\rm BH}$ is higher than $2\%\dot{M}_{\rm Edd}$, because some inflowing gas may be depleted via star formation or circularize and fall in at a slower rate. It is our future plan to improve the sub-grid physics adopted here.

Finally, an issue worth noting is whether the wind driven from the accretion flows is able to reach large scales and be injected to our simulation region. \citet{2020ApJ...890...80C} and Cui \& Yuan (2020) study the large-scale dynamics of wind launched from hot accretion flow and thin disks in the cases of without and with magnetic field via analytical and numerical methods. They find that even when magnetic field is not included, wind in the hot mode can reach very large scales while wind launched by thin disk stops at a smaller distance due to its smaller Bernoulli parameter. They do not consider the radiation pressure, which might be the essential mechanism to drive the wind in the cold mode \citep{2003MNRAS.345..657K,2015ARA&A..53..115K,2018MNRAS.473.4197C}, and the wind terminal velocity will be higher with the assistance of magnetic fields \citep{2020ApJ...890...81C}. Since in general the wind properties are function of radius, we have made sure that the properties adopted in the present work is suitable for the  radius of our inner boundary of our simulation, $\sim r_{\rm in}$.

\subsection{Simulation Setup}
We use the parallel {\tt ZEUS-MP/2} code \citep{2006ApJS..165..188H} to study an isolated elliptical galaxy by adopting two-dimensional axisymmetric spherical coordinates. The simulations start at 2~Gyrs after the Big Bang to avoid the early stage of galaxy formation when the major merger plays a dominant role, since we only focus on an isolated elliptical galaxy. The initial conditions for the ISM are set by the very low density gas at the local thermalization temperature. The distribution of the gas is not important since the mass is dominated by the stellar winds as simulations begin. The mass injection of stellar winds reaches the peak at the beginning of the simulation, then decreases gently till $z=0$, which lasts about 12~Gyrs. The mesh in the polar direction is divided homogeneously by 30 grids. In the radial direction the simulation region is resolved by 120 grids to cover the range of $2.5~\mathrm{pc}-250~\mathrm{kpc}$, with the finest resolution up to $\sim0.3~\mathrm{pc}$ by adopting a logarithmic mesh. 

\begin{figure}
	\centering
	\includegraphics[width=\columnwidth]{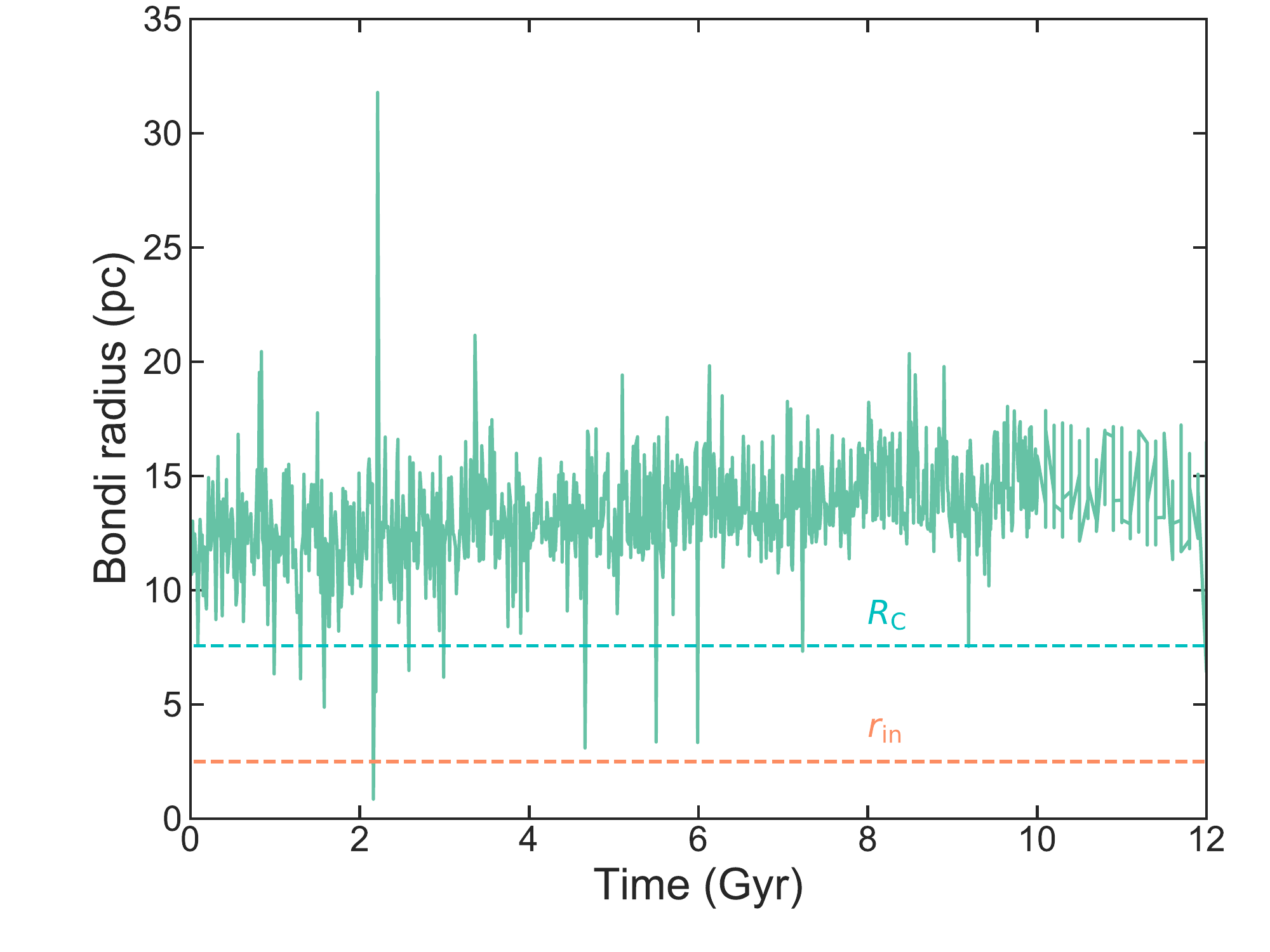}
   \caption{The Bondi radius as a function of time. The Bondi radius is calculated through the mass flux-weighted average of various phases of the gas, and only inflowing material is taken into account. The cyan and orange dashed lines represent the Compton radius for $T_\mathrm{C}=10^8~$K and the radius of the inner boundary $r_\mathrm{in}$.}
   \label{fig:bondi}
\end{figure}

How is this inner boundary  compared to the scale of accretion flow, i.e., the Bondi radius? In our simulation, the gas is in general multi-phase, and consists of both inflowing and outflowing material. Only inflowing gas is taken into account when we estimate the Bondi radius since only these gas contribute to the accretion. For inflowing gas, each phase has its corresponding Bondi radius. We have calculated the mass flux-weighted value of Bondi radius by including all phases close to our inner boundary. The calculated Bondi radius as a function of time is shown in Figure~\ref{fig:bondi}. As we can see from the figure, the Bondi radius is about 5 times larger than the radius of our inner boundary.  We have also calculated  the Compton radius, which is defined by
\begin{equation}
R_\mathrm{C}=\frac{2GM_\mathrm{BH}\mu m_\mathrm{p}}{3k_\mathrm{B}T_\mathrm{C}}\simeq3.5\mu\frac{M_\mathrm{BH}}{10^9~M_\odot}\frac{10^8~\mathrm{K}}{T_\mathrm{C}}~\mathrm{pc}.
\end{equation}
It is larger than the inner boundary radius for $T_\mathrm{C}=10^8~$K. This indicates that the Compton effect mainly plays a role of cooling inside the inner boundary. We carve a narrow range of $\theta$ ($\sim9\degr$) at each pole to avoid the singularity there and also adopt the ``outflow boundary condition'' for polar boundaries. The AGN sub-grid model works as follows. After measuring the accretion rate at the inner boundary, we can calculate the BH accretion rate and wind properties by adopting the formulas in the corresponding mode. Then the BH mass is updated and the wind is injected in the inner grid within the opening angle with mass and momentum conservation. A temperature floor of $10^4~\mathrm{K}$ is set in the cooling functions due to the considerations of spatial resolution in our simulations.

\begin{table*}
   \centering
   \caption{Summary of the surveyed parameters.}
   \label{tab:param}
   \begin{threeparttable}
   \newcolumntype{Y}{>{\centering\arraybackslash}X}
   \begin{tabularx}{1.01\textwidth}{cYYYYYYY}
       \toprule
       \multirow{2}*{Run} & \multicolumn{3}{c}{Hot Mode} & \multicolumn{3}{c}{Cold Mode} & \\
       \cmidrule(lr){2-4} \cmidrule(lr){5-7}
       & $v_\mathrm{W,H}$\tnote{*} & $\dot{M}_\mathrm{W,H}$ & $\varepsilon_\mathrm{hot}$  & $v_\mathrm{W,C}$ & $\dot{M}_\mathrm{W,C}$ & $\varepsilon_\mathrm{cold}$ &  $M_\mathrm{BH,i}$ (M$_\odot$)\\
       \toprule
       Fidu & x & x & $\delta=0.1$ & x & x & 0.1 & $1.8\times10^9$ \\
       \midrule
       HotWindVel & 3x & &  &  &  &   &  \\
       HotWindFluxHigh & & $\dot{M}_\mathrm{W,SANE98}$ &  &  &  &   &  \\
       HotWindFluxLow & & $\dot{M}_\mathrm{W,MAD00}$&  &  &  &   &  \\
       HotRad & & & $\delta=0.5$ & & & & \\
       \midrule
       ColdWindVelHigh & & & & 3x & & & \\
       ColdWindVelLow & & & & 0.33x & & & \\
       ColdWindFluxHigh & & & & & 10x& & \\
       ColdWindFluxLow & & & & & 0.1x& & \\
       ColdRadHigh & & & & & & 0.3 & \\
       ColdRadLow & & & & & & 0.057 & \\
       \midrule
       BHmass & & & & & & & $2.7\times10^8$\\
       \bottomrule
   \end{tabularx}
   \begin{tablenotes}
   \item[*] Meaning of symbols. $v_\mathrm{W,H}$: wind velocity in the hot mode; $\dot{M}_\mathrm{W,H}$: wind mass flux in the hot mode; $\varepsilon_\mathrm{hot}$: radiative efficiency of the accretion flow in the hot mode;  $v_\mathrm{W,C}$: wind velocity in the cold mode; $\dot{M}_\mathrm{W,C}$: wind mass flux in the cold mode; $\varepsilon_\mathrm{cold}$: radiative efficiency of the accretion disk in the cold mode; $M_\mathrm{BH,i}$: initial BH mass; $\delta$: the fraction of viscously dissipated energy that directly heats electron.
   \end{tablenotes}
   \end{threeparttable}
\end{table*}

\section{Model Parameter Explorations}
\label{sec:parameter}
In this section, we introduce the explorations of our model parameters. We perform surveys on wind and radiation of the AGN in both hot and cold modes. To avoid confusion, each model of wind and radiation alters one parameter while keeping other parameters unchanged to control variables. Apart from that, we explore the case of a lower initial BH mass. Note that in this case the Eddington accretion rate decreases due to the lower BH mass. Therefore, the same mass flux of accretion flows yield higher Eddington ratio. All the surveyed parameters are summarized in Table~\ref{tab:param}. These parameters are explored within the constraints imposed by observations or theories, as detailed below.

\begin{figure}
	\centering
	\includegraphics[width=\columnwidth]{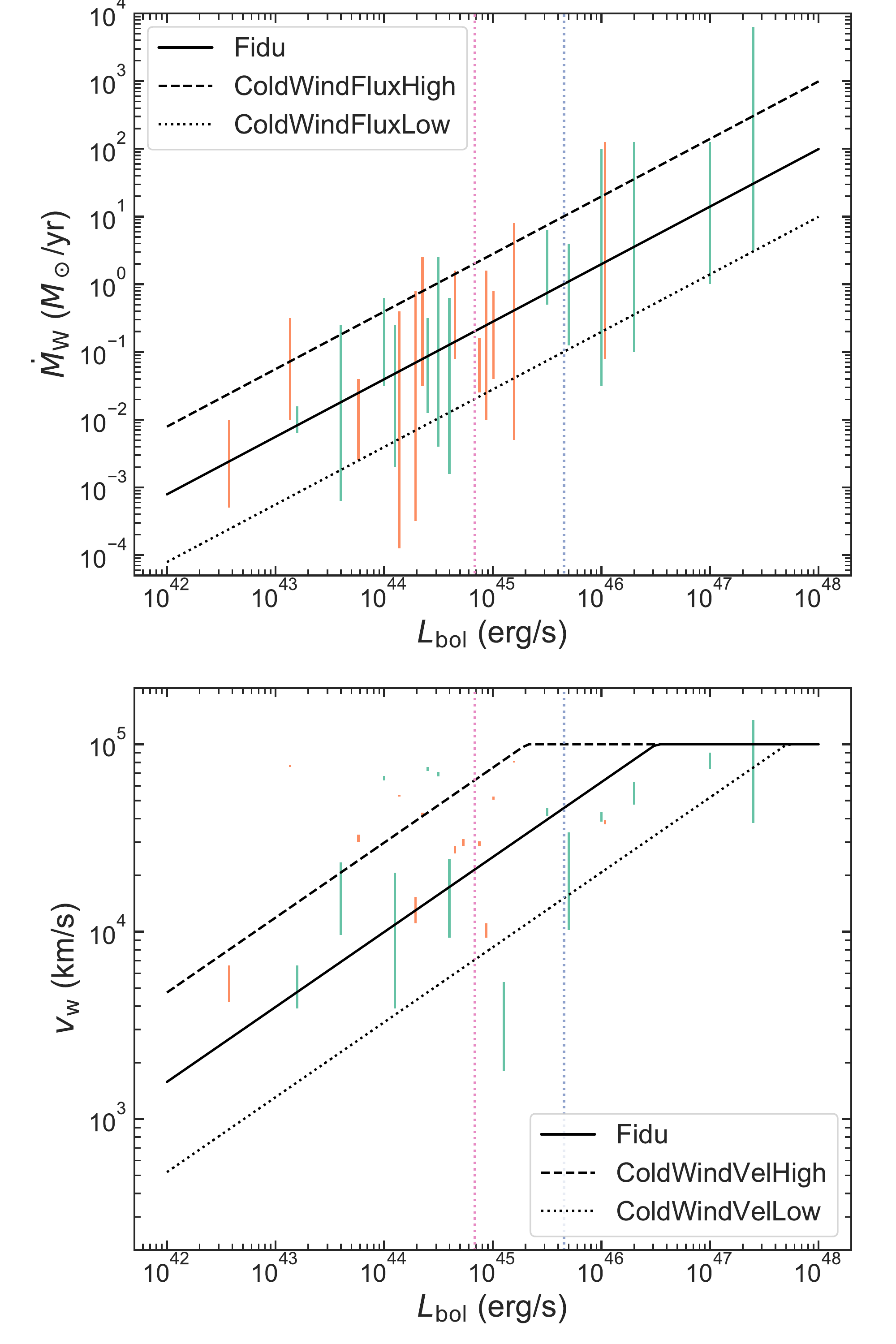}
   \caption{The mass flux (top) and velocity (bottom) of wind in the cold mode as a function of AGN  bolometric luminosity $L_\mathrm{bol}$. The \textbf{green} and orange data points are taken from  the \textit{Suzaku} sample in \citet{2015MNRAS.451.4169G} and the \textit{XMM-Newton} sample in \citet{2012MNRAS.422L...1T} respectively.  The solid lines are the best fittings to the green data points given in \citet{2015MNRAS.451.4169G}. The dashed and dotted lines vary the best fittings by a factor of ten (top) and three (bottom), representing the ColdWindFluxHigh/Low models and the ColdWindVelHigh/Low models, respectively.  The magenta and gray dotted lines represent the critical luminosity ($2\%~L_\mathrm{Edd}$) for the Fidu and BHmass models.  }
   \label{fig:ColdWind}
\end{figure}

\begin{figure}
	\centering
	\includegraphics[width=\columnwidth]{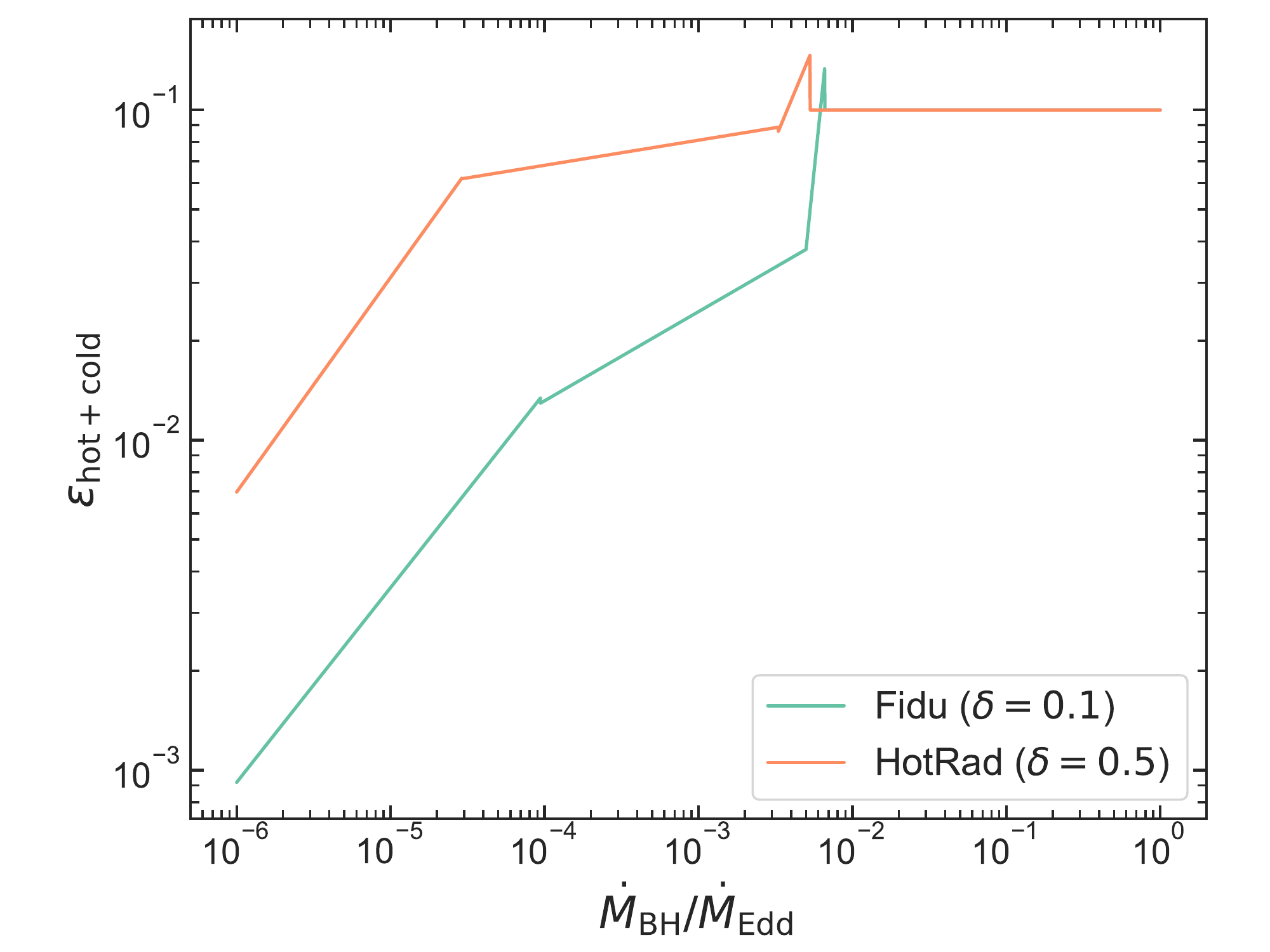}
   \caption{Radiative efficiencies corresponding to two values of  the hot accretion flow parameter $\delta$ as a function of BH accretion rate. The larger $\delta$ denoted by the orange line is adopted in the HotRad model.}
   \label{fig:Radiation}
\end{figure}
\subsection{Wind}
\label{subsec:wind}

In the hot mode due to the scarce of observational data, we model the wind properties by adopting simulation results of \citet{2015ApJ...804..101Y}. This work only deals with non-spin black hole and normal magnetic field (SANE). On the basis of \citet{2015ApJ...804..101Y}, Yang et al. (2020, in prep.) study the effects of magnetic field and BH spin on wind properties by performing new GRMHD simulations. Their results suggest that the magnetic field and BH spin can affect the wind properties significantly. Wind velocity in the case of strong magnetic field (MAD00, here the number denotes the spin of BH) is about three times higher than that in its weak magnetic field counterpart (SANE00) obtained in \citetalias{Yuan:2018ec}. HotWindVel is the model to explore this high velocity case. The mass flux of wind of MAD00 is given by
\begin{equation}
\dot{M}_\mathrm{W,MAD00}=\dot{M}_\mathrm{BH}\left(\frac{r_\mathrm{tr}}{55~r_\mathrm{s}}\right)^{1.54}.
\end{equation}
This is several times lower than that of SANE00 when the mass accretion rate approaches $10^{-2}~\dot{M}_\mathrm{Edd}$. On the other hand, the mass flux of the high BH spin model SANE98 given by
\begin{equation}
\dot{M}_\mathrm{W,SANE98}=\dot{M}_\mathrm{BH}\left(\frac{r_\mathrm{tr}}{6~r_\mathrm{s}}\right)^{0.95}
\end{equation}
is three times higher than that of SANE00. It is worth noting that when $\dot{M}(r_\mathrm{in})$ is small, all three models yield $\dot{M}_\mathrm{W}$ close to $\dot{M}(r_\mathrm{in})$, because $\dot{M}_\mathrm{BH}$ of all three models is orders of magnitude smaller than $\dot{M}_\mathrm{W}$. We perform the HotWindFluxLow and HotWindFluxHigh models to simulate these two cases of MAD00 and SANE98, respectively. The mass flux of the HotWindVel model and the wind velocity of the HotWindFluxHigh/Low model remain unchanged under the same $\dot{M}(r_\mathrm{in})$. This indicates the lower wind density of the HotWindVel model and higher wind density of the HotWindFluxHigh/Low model given the definition of mass flux
\begin{equation}
\dot{M}_\mathrm{W}=\rho_\mathrm{W} v_\mathrm{W} A,
\end{equation}
where $A$ is the constant area that the wind blows out of the inner boundary.

In the cold mode we estimate wind properties through observations on broad absorption line (BAL) outflows \citep{1999ApJ...516...27A,2010A&A...521A..57T,2012MNRAS.422L...1T,2013MNRAS.436.3286A,Gofford2013,2015MNRAS.451.4169G}. \citetalias{Yuan:2018ec} adopts the relation between wind properties and bolometric luminosity from \citet{2015MNRAS.451.4169G}. Figure~\ref{fig:ColdWind} compares this relationship with the observational data on UFOs (ultra-fast outflows). In addition to the data from \citet{2015MNRAS.451.4169G}, the data from \citet{2012MNRAS.422L...1T} is also shown in the figure. It is worth noting that because the wind injection radius, i.e., the inner boundary, in our sub-grid model is within the Bondi radius, we only choose observations of UFOs that are on sub-pc scales. From Figure~\ref{fig:ColdWind}, it is easy to see that the uncertainties of the wind flux and velocity are considerable. Therefore, we perform four models ColdWindFluxHigh/Low, ColdWindVelHigh/Low, trying to cover the scatters in the observations. The flux-variant models vary the mass flux by an order of magnitude and the velocity-variant models alter the wind velocity by a factor of three. Note that an upper limit of $10^5$~km/s is applied in all the models. Similarly, the mass flux of the ColdWindVelHigh/Low model and wind velocity of the ColdWindFluxHigh/Low model remain identical.

\subsection{Radiation}
In addition to wind properties, the effect of radiation is studied as well in both hot and cold modes. The HotRad model studies the effect of radiation in the hot mode. As shown in Section~\ref{sec:agnfb}, the radiative efficiency is a function of $\delta$, which denotes the fraction of viscously dissipated energy that directly heats electrons. The value of $\delta$ is constrained to be  $\delta\sim0.1-0.5$ \citep{2014ARA&A..52..529Y},  stronger radiation corresponding to larger $\delta=0.5$ is considered in the HotRad model. Figure~\ref{fig:Radiation} shows the radiative efficiencies corresponding to two different $\delta$ as a functions of BH accretion rate. The radiative efficiency corresponding to $\delta=0.5$ can be an order of magnitude higher than that of $\delta=0.1$ when the BH accretion rate is low. The radiative efficiency of the HotRad model in the cold mode remains  same.

In the cold mode, the ColdRadHigh model explores the high spin of the BH with radiative efficiency $\varepsilon_\mathrm{cold}=0.3$, while the ColdRadLow model investigates the zero spin of BH, of which the radiative efficiency  $\varepsilon_\mathrm{cold}=0.057$. In order to study the effects of radiation in the cold mode, we keep the the radiative efficiency in the hot mode identical to that of the Fidu model. Moreover, in order to study radiation, the wind properties remain unchanged under the same $\dot{M}(r_\mathrm{in})$.

\subsection{Initial BH mass}
Other than wind and radiation properties, we explore the effect of the initial BH mass using the BHmass model. The initial BH mass adopted in \citetalias{Yuan:2018ec} is $M_\mathrm{BH,i}=1.8\times10^{9}~M_\odot$ for $M_\star=3\times10^{11}~M_\odot$ \citep{Kormendy_Coevolution_2013}, but recent works suggest the discrepancy of BH mass measurement between AGN hosts and ellipticals/classical bulges \citep{2014ApJ...789...17H,2015ApJ...813...82R,2016MNRAS.460.3119S,2019MNRAS.485.1278S}. \citet{2016MNRAS.460.3119S} use Monte Carlo simulations to illustrate the selection bias in local, dynamically measured BH samples of ellipticals/classical bulges, in which only the massive BHs are measured. Therefore, the BH mass in \citet{Kormendy_Coevolution_2013} may be overestimated. In their model the BH mass is $M_\mathrm{BH,i}=2.7\times10^{8}~M_\odot$ for $M_\star=3\times10^{11}~M_\odot$. The BHmass model studies this possibility, with the initial BH mass in this model being six times lower than that in \citetalias{Yuan:2018ec}.

\begin{figure*}
	\centering
	\includegraphics[width=\textwidth]{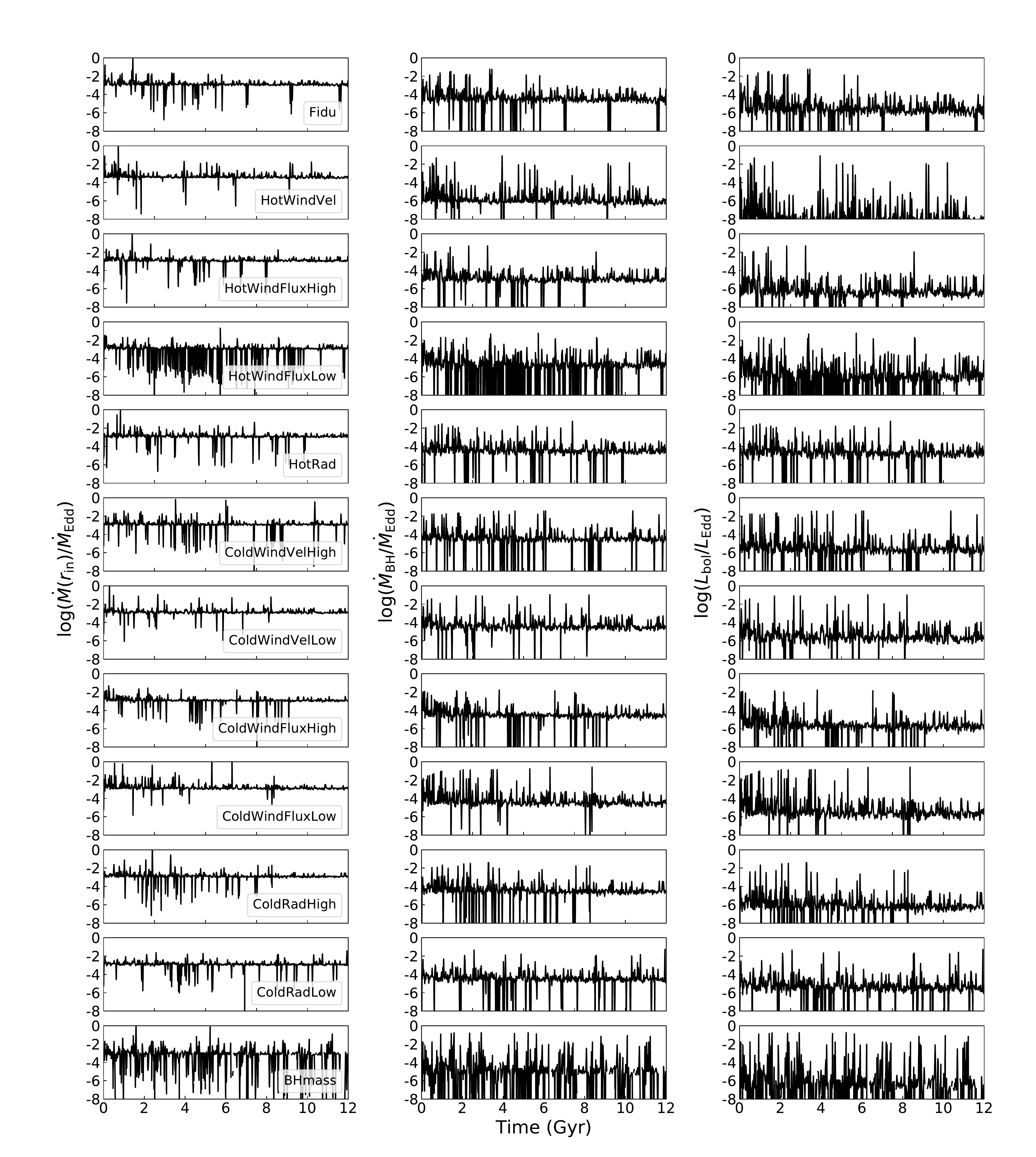}
   \caption{Evolution of AGNs for various models: [left] mass accretion rate at the inner boundary, [middle] accretion rate at the BH horizon, [right] AGN light curves. All these three quantities are scaled by Eddington limits.}
   \label{fig:lc}
\end{figure*}

\begin{figure}
	\centering
	\includegraphics[width=\columnwidth]{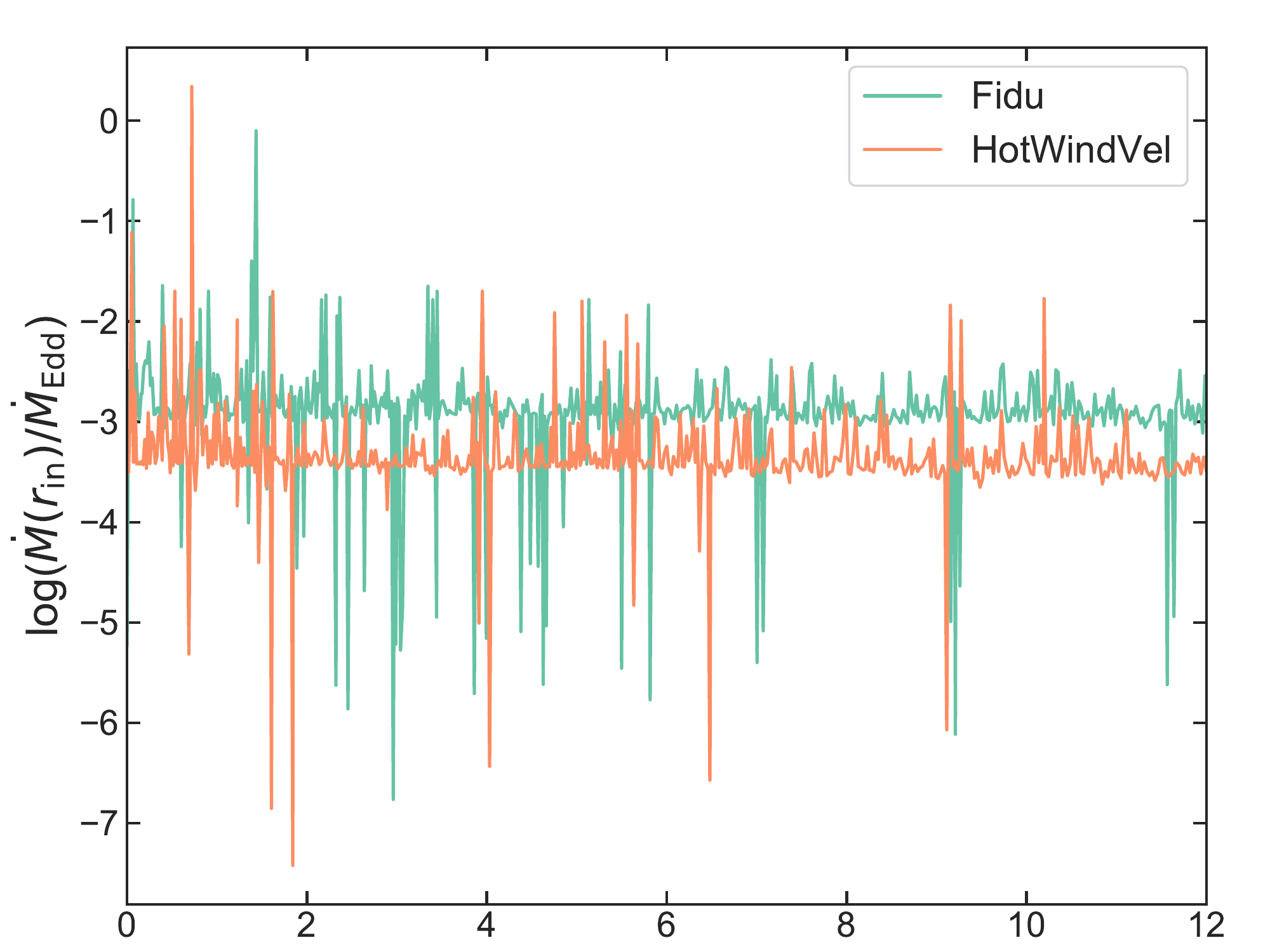}
   \caption{Same as the first column of Figure~\ref{fig:lc}, but we only show the Fidu and HotWindVel models here for clear comparison. The typical accretion rate of the HotWindVel model is lower than that of the Fidu model by a factor of three.}
   \label{fig:mdotin}
\end{figure}

\section{Results}
\label{sec:results}

In this section, we present the results of our parameter explorations. We essentially investigate the effects of parameters in four major domains of galaxy evolution: AGN luminosity, BH mass growth, star formation, and the AGN duty cycle. The fiducial model is essentially identical to the fullFB model in \citetalias{Yuan:2018ec} but with two improvements as discussed in \citet{2019ApJ...885...16Y}. One is the calculation of star formation as described in Section~\ref{subsec:sf}, the other is the correction of a bug in computing the energy flux of the wind. This bug caused a lower energy flux of the wind in the hot mode while a higher energy flux in the cold mode.

\subsection{AGN luminosity}
\label{subsec:lum}

Figure~\ref{fig:lc} shows the evolution of the accretion rate at the inner boundary $\dot{M}(r_\mathrm{in})$,  BH accretion rate $\dot{M}_\mathrm{BH}$, and bolometric luminosity $L_\mathrm{bol}$ for all models. We note that the time interval of two adjacent data points is $\sim1~$Myr for clarity, thus some outbursts are filtered out in this case.
For the fiducial model, $\dot{M}(r_\mathrm{in})$ oscillates around $\sim10^{-3}~\dot{M}_\mathrm{Edd}$. The value of this ``baseline accretion rate'' is determined by the momentum balance between the AGN wind and the inflow at the inner boundary of our simulation, reflecting the characteristic of the hot mode in our AGN sub-grid model. When the accretion rate decreases, both the mass flux and the velocity of the wind become lower and yield a lower momentum of the wind. According to our sub-grid model, the decrease of wind momentum is more rapid than that of the inflow, thus making the accretion rate increase to reach momentum balance. Oppositely, when the accretion raises, the wind momentum grows faster than the momentum of inflow, thus making the accretion rate decrease. Therefore, we can see small oscillations of $\dot{M}(r_\mathrm{in})$ at $\sim10^{-3}~\dot{M}_\mathrm{Edd}$. However, if the inflow is massive enough to overcome the momentum of wind, such as driven by the cold clumps, the accretion rate is able to enter the cold mode and trigger a strong AGN wind and radiation of the cold mode. The wind and radiation push the accreting gas outwards so the mass accretion rate at the Bondi radius drops drastically. In addition, they interact with the ISM of the nuclear region and heat the surrounding gas. The subsequent galactic wind can reach several kpc scale, but most of the wind can not break out of the halo. The compression of the kicked gas is likely to form the cold clumps once more, yet in a moderate way because of the lower density of the ISM. These cold clumps fall back to the centre and initiate another episode of the AGN outburst. Eventually, the surrounding gas is almost cleared out and the AGN returns back to be quiescent. Because the mass supply from the stellar wind decreases with time, the massive inflow characterized by the peaks of accretion rate occurs more frequently at the early stage of the evolution due to abundant supplies from the stellar wind. Actually, the accretion rate stays in the hot mode after $\sim6$~Gyr. However, the accretion rate for the momentum balance maintains the same level throughout the evolution, keeping the typical mass accretion rate unchanged.

As shown in the middle panel of Figure~\ref{fig:lc}, the BH accretion rate oscillates around $10^{-5}~\dot{M}_\mathrm{Edd}$, and the amplitude of the oscillation is larger than that of  $\dot{M}(r_\mathrm{in})$. Given that $\dot{M}(r_\mathrm{in})\sim10^{-3}~\dot{M}_\mathrm{Edd}$, this implies that $99\%$ of the inflows are restored to the simulation region by the strong wind. The right panel shows the light curve of the AGN $L_\mathrm{bol}=\varepsilon\dot{M}_\mathrm{BH}c^2$. Overall, we can see that the galaxy spends most of its time in the low-luminosity phase with the bolometric luminosity lying in the range $10^{-6}-10^{-4}~L_\mathrm{Edd}$, which is consistent with observations that a median Eddington ratio of $\sim10^{-5}$ is found for ellipticals \citep{2009ApJ...699..626H}. Our typical luminosity of Fidu is 1-2 orders of magnitude lower than that of the fullFB model in \citetalias{Yuan:2018ec}, which is $\sim10^{-4}~L_\mathrm{Edd}$. This is caused by the correction of the bug that under-produced the wind power in the hot mode.

As the parameters vary, Figure~\ref{fig:lc} shows that $\dot{M}(r_\mathrm{in})$ varies the least among the three physical quantities, which reflects the nonlinear characteristics in our AGN sub-grid models. Here, variations are measured through typical values the curves fluctuate around and the frequency of strong oscillations, such as outbursts and sudden drops to very low accretion rate or luminosity. For the HotWindVel model, $\dot{M}(r_\mathrm{in})$ oscillates around a value a factor of three smaller than that of the Fidu model (Figure~\ref{fig:mdotin}), resulting in orders of magnitude lower BH accretion rate and bolometric luminosity \footnote{To be specific, the time-averaged $\dot{M}_\mathrm{Bondi}$ is $1.5\times10^{-3}~\dot{M}_\mathrm{Edd}$ for the Fidu model, and is $5\times10^{-4}~\dot{M}_\mathrm{Edd}$ for the HotWindVel model.}. As mentioned above, the oscillations are caused by the momentum balance between the AGN wind and the inflow. When the wind velocity is three times higher,   it requires $\dot{M}(r_\mathrm{in})$ three times lower compared to the Fidu model to achieve a new momentum balance between the inflow and wind at $r_\mathrm{in}$. This is because, a three-times lower $\dot{M}(r_\mathrm{in})$  will result in both the wind flux and velocity three times lower, while the higher wind velocity in the HotWindVel model will compensate for the decrease of velocity. So the overall  decrease of momentum of wind is three times, which balances with the decrease of inflow.

We find from Figure~\ref{fig:lc} that the typical accretion rates of the HotWindFluxHigh and HotWindFluxLow models vary little compared to the Fidu model. This is because, the mass fluxes of wind in these two models differ from the Fidu model only when the accretion rate is close to $10^{-2}\dot{M}_{\rm Edd}$, as we emphasize in
Section~\ref{subsec:wind}.  When $\dot{M}(r_\mathrm{in}) \la 10^{-3}~\dot{M}_\mathrm{Edd}$, the mass fluxes of the hot wind for all the models are similar, and are roughly equal to  $\dot{M}(r_\mathrm{in})$. Yet the HotWindFluxHigh model has fewer outbursts, while for the HotWindFluxLow model, more frequent outbursts can be seen during its violent evolution. Such a difference is caused by the different mass fluxes of hot wind when $\dot{M}(r_\mathrm{in})$ approaches $10^{-2}~\dot{M}_\mathrm{Edd}$. For the HotWindFluxHigh model, the wind is stronger than that of the Fidu model; thus with increasing accretion rate, the accretion mode is more likely to stay in the hot mode. For the HotWindFluxLow model, the wind is weaker, thus with the increase of accretion rate, it is easier for the accretion to enter the cold mode. Since the wind is stronger in the cold mode, it can cause larger oscillations characterized by the outbursts and suppressions of accretion rate and luminosity.

\begin{figure*}
	\centering
	\includegraphics[width=\textwidth]{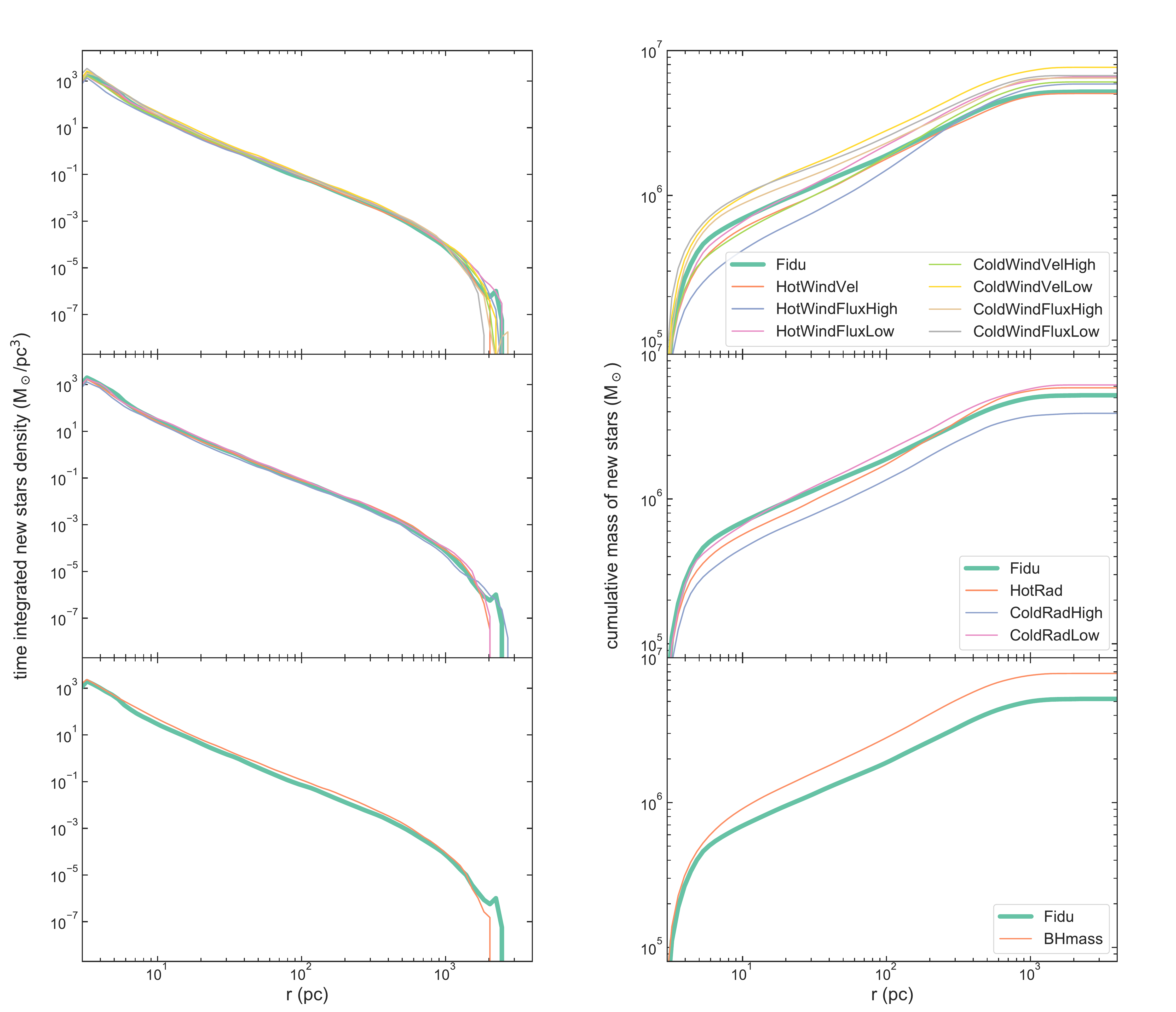}
   \caption{The distribution of newly born stars as a function of radius for various models \textbf{at the end of the simulations}: [left] new stars density, [right] cumulative mass of new stars.}
   \label{fig:ns}
\end{figure*}

The HotRad model shows similar $\dot{M}(r_\mathrm{in})$ and $\dot{M}_\mathrm{BH}$ compared to the Fidu model, but the typical $L_\mathrm{bol}$  is an order of magnitude higher due to the higher radiative efficiency. The results of the four models concerning properties of cold wind exhibit less variation than those of the Fidu model. Comparing the High-suffix models (i.e., ColdWindVelHigh and ColdWindFluxHigh) to the Low-suffix counterparts (i.e., ColdWindVelLow and ColdWindFluxLow) of the cold mode, we can see from the figure that the strong suppression of the accretion rate appears more frequently and has larger amplitudes due to stronger wind feedback. But the effect of radiative efficiency in the cold mode is less obvious.  Finally, the accretion rate of the BHmass model shows strong oscillations, which is similar to that of the HotWindFluxLow model. This is because the Eddington accretion rate is lower due to the lower BH mass. Therefore it becomes easier for the accretion to enter into the cold mode given the same amount of mass supply from the stellar wind, which produces a stronger wind.
l, we find that the galaxy spends more time in the low-luminosity phase with increasing mass flux and velocity of the hot wind as shown by HotWindFluxLow, HotWindFluxHigh, and HotWindVel. Correspondingly the BH growth and the datum line luminosity decline as we mentioned above. Yet all of these three models spend less time in the cold mode. We suppose that the reason of HotWindFluxLow is unphysical as we have discussed already. Furthermore, all of these three models differ from the observations by \citet{2009ApJ...699..626H}, indicating these cases are a rarity. Although the galaxies of HotWindVel and HotWindFluxHigh stay longer time in the hot mode, the emitted energy fraction of the cold mode is higher than that of Fidu. This is because the energy/time shown here is the proportion of the total energy/time. Owing to the fact that they spend more time during the lower luminosity phase, in other words, the datum line luminosity is lower, the total energy they emit is lower than that of Fidu, which causes their cumulative energy above $2\%~L_\mathrm{Edd}$ relatively higher. Overall, the variation of the cumulative time and energy above $2\%~L_\mathrm{Edd}$ is within a factor of 2.

\subsection{BH mass growth}
\label{subsec:bh}
The BH mass growth for various models are listed in the second column of Table~\ref{tab:effects}. Generally all the models yield mass growth of less than $4\%$ of the initial BH mass, suggesting that the AGNs are quiescent overall regardless of the parameters we explore.  Moreover, the models with stronger wind or radiation have lower BH mass growth, suggesting that both AGN wind and radiation have negative effects on the BH mass growth, which is easy to understand.

Specifically, as a result of increasing the wind velocity by a factor of three, the HotWindVel model has lower BH mass growth due to its lower typical BH accretion rate in the hot mode. The HotWindFluxHigh model reduces the BH mass growth by over a factor of two, while the HotWindFluxLow model has higher BH mass growth due to the lower mass flux of the wind. In the cold mode, while the BHs have over $50\%$ higher mass growth by reducing velocity and mass flux of the cold mode as shown by the ColdWindVelLow and ColdWindFluxLow models, respectively, the BH mass varies little when increasing the velocity and mass flux by the same magnitude, as shown by the ColdWindVelHigh and ColdWindFluxHigh models. This is caused by the small fraction of BH mass growth in the cold mode. Since the growth of BH mass is already dominated by the accretion in the hot mode, stronger feedback in the cold mode reduces little BH growth. However, weaker feedback in the cold mode can increase the accretion during the cold mode significantly, thus resulting in more substantial BH mass variations.

The models with stronger radiation either in the hot mode or the cold mode show negative effects on the BH accretion. Weaker radiation in the cold mode represented by the ColdRadLow produces higher BH mass growth. But the effect of radiation is smaller than the wind,  consistent with \citetalias{Yuan:2018ec}. However, we should be cautious in drawing the conclusion that wind is more important than radiation in controlling the BH accretion, since we have not considered dust in our simulations, which might play an important role in the radiation feedback processes.

Finally, the BHmass model shows much smaller growth of the BH mass than that of the Fidu model. The reasons are twofold. One is that the lower BH mass provides a shallower gravitational potential that reduces the accretion. The other is that the Eddington accretion rate is lower for the BHmass model. Consequently, the accretion is easier to enter the cold mode given the same amount of mass supply from stellar wind, while the cold mode has stronger wind and radiation feedback. However, the percentage of the mass growth of the BHmass model is higher than that of the Fidu model, caused by the higher ratio of stellar mass to the initial BH mass.

\begin{table}
   \centering
   \begin{adjustbox}{max width=0.48\textwidth}
   \begin{threeparttable}
   \caption{Effects of the surveyed parameters.}
   \label{tab:effects}
   \begin{tabular}{lcccc}
       \toprule
       \multirow{2}{*}{Run} & BH mass growth & Total star formation & Duty cycle\tnote{*}\\
        & ($\times10^7~$M$_\odot$) & ($\times10^6~$M$_\odot$)&  (\%) \\
       \midrule
       Fidu           & 3.73\tnote{**}  & 5.84 & 0.04         \\
       HotWindVel    & 0.75   & 5.76  & 0.02      \\
       HotWindFluxHigh & 1.64  & 6.70 & 0.04    \\
       HotWindFluxLow & 4.48  & 7.57   & 0.04         \\
       HotRad     & 3.68    & 6.71    & 0.04    \\
       ColdWindVelHigh  & 3.57  & 6.89  &  0.03     \\
       ColdWindVelLow  & 5.69   & 8.65  &  0.10       \\
       ColdWindFluxHigh & 3.37 & 7.31 &  0.00        \\
       ColdWindFluxLow & 6.52  & 7.68  &  0.10      \\
       ColdRadHigh  & 3.52  & 4.38 & 0.03        \\
       ColdRadLow  & 3.80  &  6.86 & 0.20      \\
       BHmass & 0.96   & 8.81  & 0.10       \\
       \bottomrule
   \end{tabular}
   \begin{tablenotes}
   \item{*} The fraction of time that galaxy spends above 0.02 $L_{\rm Edd}$.
   \item[**] The BH mass growth of the Fidu model is an order of magnitude smaller than that of the fullFB model in \citetalias{Yuan:2018ec}. It is reasonable considering an order of magnitude lower typical BH accretion rate for the Fidu model.
   \end{tablenotes}
   \end{threeparttable}
   \end{adjustbox}
\end{table}

\subsection{Star formation}
Figure~\ref{fig:ns} shows the distribution of the newly born stars for various models. We ignore the gravitational potential provided by the new formed stars and the stellar movements. The results differ significantly from \citetalias{Yuan:2018ec} because we now apply different density and temperature thresholds to star formation, i.e., only the gas with density over $1~\mathrm{cm}^{-3}$ and temperature under $4\times10^4~\mathrm{K}$ can form stars, and star formation efficiency $\eta_\mathrm{SF}$ is reduced by an order of magnitude.  Generally, we find that star formation is significantly reduced compared to that of the fullFB model in \citetalias{Yuan:2018ec}, and is highly concentrated inside 1~kpc. The new stars density at the end of the simulations, shown in the left panel of Figure~\ref{fig:ns}, decreases with increasing radius due to the increasing star formation timescale and decreasing gas density. The total mass of new stars is around $(4-9)\times10^6~M_\odot$ for various models, shown in the middle column of Table~\ref{tab:effects}, which accounts for less than $0.01\%$ of the total stellar mass. In fact, the total star formation is lower than the BH mass growth for all the models.

We do not find obvious correlations between the wind properties and the total star formation, as shown by the upper right panel of Figure~\ref{fig:ns}, which presents the cumulative mass of new stars integrated over time for models with various wind parameters. In the simulations, the sites of star formation are determined by the distribution of the cold and dense gas. A stronger AGN wind is able to push the gas toward larger radii, and results in the accumulation and condensation of the cold gas there. Thus higher wind power increases star formation at large radii while decreasing it at small radii. On the other hand,  wind feedback does strongly suppress the star formation, as shown by, e.g., the right plot of Fig. 8 in \citetalias{Yuan:2018ec}.  The absence of the correlation found here is because the  range of wind parameters we have explored is relatively too small.

In the center right panel, the HotRad model shows that higher radiative efficiency of the hot mode yields lower star formation inside 100~pc, caused perhaps by stronger radiative heating and radiation pressure;  while at large radii, star formation is higher, which is perhaps because the radiation pressure pushes the gas from small radii to large radii, similar to the role played by wind.  For radiation in the cold mode, however, stronger radiation suppresses the star formation at both small and large radius, which is likely because the radiative heating is very strong in the cold mode. The bottom right panel shows that the lower BH mass of the BHmass model produces higher star formation overall.

\begin{figure}
	\centering
	\includegraphics[width=0.5\textwidth]{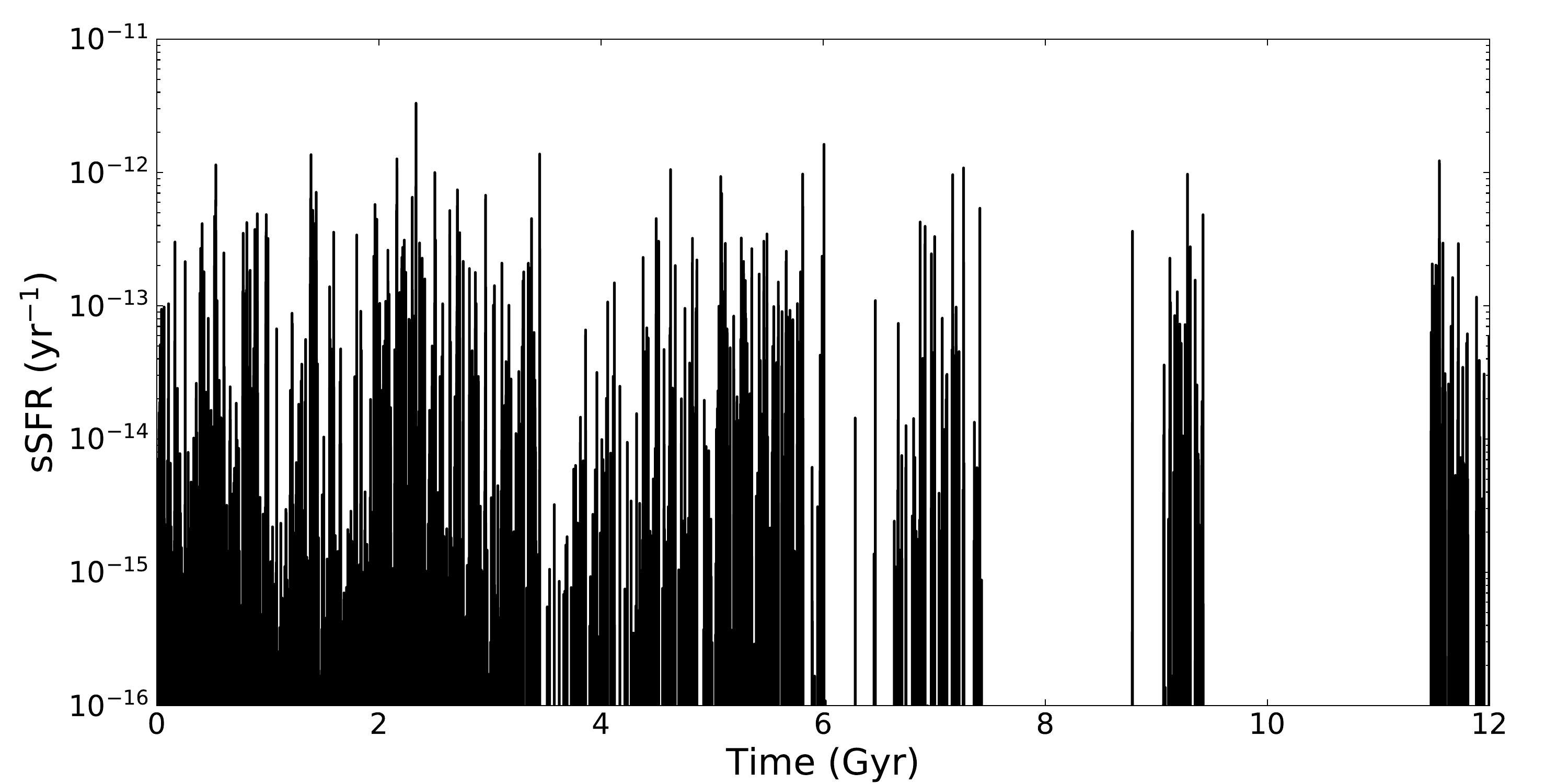}
   \caption{The time evolution of specific star formation rate for the Fidu model.}
   \label{fig:ssfr}
\end{figure}

Apart from the spatial properties of star formation, another major difference from \citetalias{Yuan:2018ec} is the temporal evolution of the specific star formation rate (sSFR). Figure~\ref{fig:ssfr} shows the evolution of the sSFR for the Fidu model as an example. Throughout the lifetime of the galaxy, the sSFR is weak for most of the time with the time-averaged value of $10^{-15}~\mathrm{yr}^{-1}$. The quiescence of the galaxy, however, is punctuated by sporadic and short episodes of strong star formation outbursts that can reach the sSFR of $10^{-12}~\mathrm{yr}^{-1}$ especially at the early stage. This star formation history is strongly correlated with the AGN luminosity light curve. We find the concurrent outbursts in the star formation and the AGN light curve, and the same quiescent time during 7-9~Gyr and 9-11~Gyr. This indicates the tight link between star formation and BH accretion. As mentioned before, the star formation takes place in the cold, dense clumps. If these clumps are not consumed by star formation before they fall into the central BH, a strong AGN outburst is expected afterwards. Both processes are driven by cooling flow infall of gas.

\begin{figure*}
	\centering
	\includegraphics[width=\textwidth]{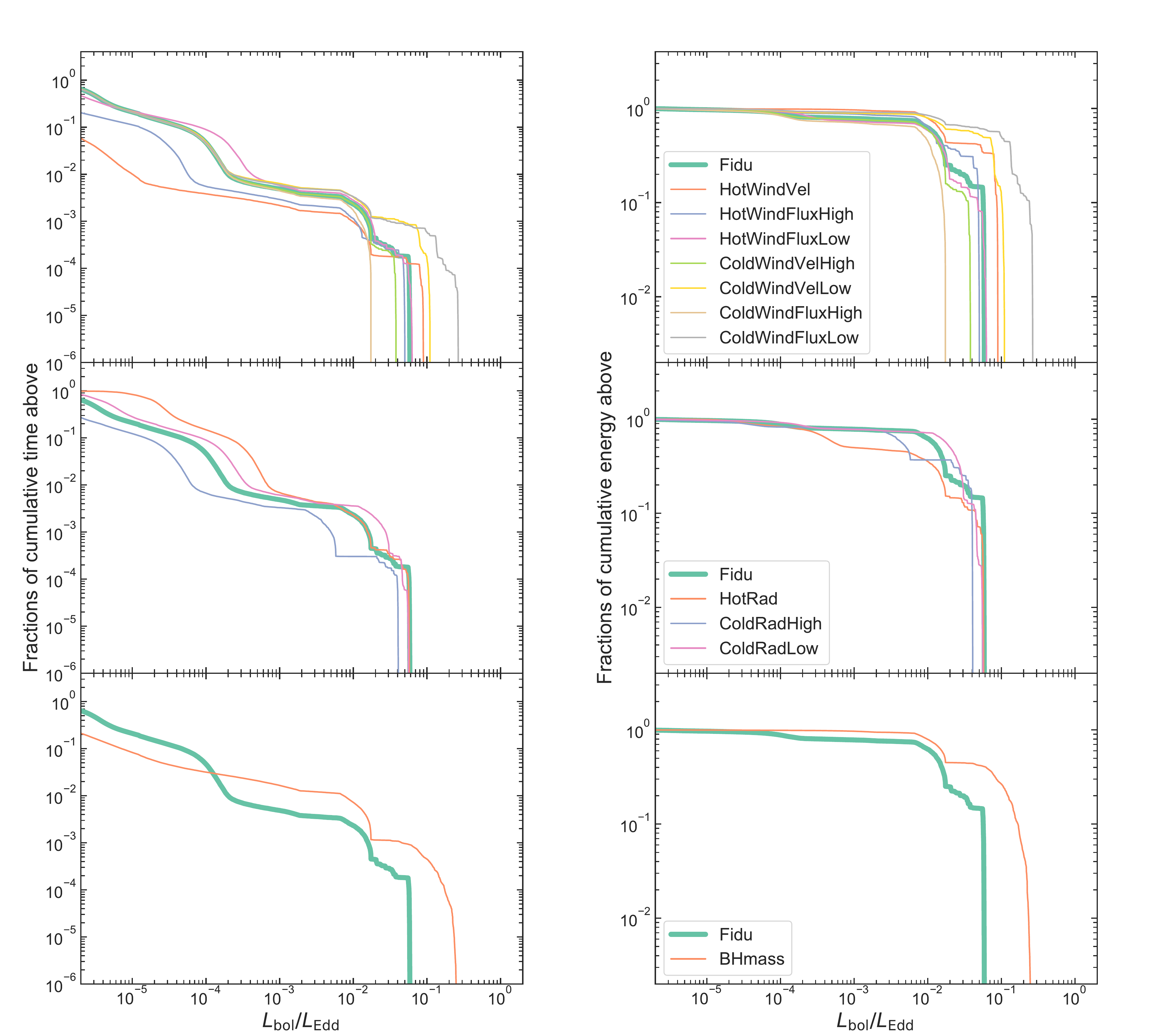}
   \caption{Fractions of the total time spent  [left] and energy emitted [right] above the values at given Eddington ratios of the central AGN  for various models.}
   \label{fig:dc}
\end{figure*}

\subsection{AGN Duty cycle}
The duty cycles for different models are shown in the left panel of Figure~\ref{fig:dc}. Each line represents the fractions of time that the galaxy spends above a given Eddington ratio.
For the Fidu model, the AGN spends over $99\%$ of its evolution time with the Eddington ratio below $2\times10^{-4}$. Comparing to $80\%$ of the total time in the fullFB model of \citetalias{Yuan:2018ec}, the AGN of the Fidu model spends more time in the low-luminosity phase, which is consistent with the lower typical AGN luminosity mentioned in Section~\ref{subsec:lum}.

In general, wind and radiation of the hot mode exert more remarkable influences on the cumulative time during the low-luminosity phase, while wind and radiation of the cold mode have larger effects above $10^{-2}~L_\mathrm{bol}/L_\mathrm{Edd}$. Specifically, we  can see from the top left panel of Figure~\ref{fig:dc} that the AGN spends more time below the Eddington ratio of $10^{-4}~L_\mathrm{bol}/L_\mathrm{Edd}$ as the velocity or mass flux of wind in the hot mode increases. For the Eddington ratio above $10^{-2}~L_\mathrm{bol}/L_\mathrm{Edd}$, however, the AGN spends less time with increasing velocity or mass flux of wind in the cold mode. Particularly, the AGN of the ColdWindFluxHigh model spends little time in the cold mode throughout its evolution. This suggests that AGN wind of both hot and cold modes has negative effects on the AGN duty cycle, which is consistent with the findings that the AGN wind suppresses the BH accretion in Section~\ref{subsec:bh}.

The center left panel of Figure~\ref{fig:dc} shows the models with regard to radiation. The AGN in the HotRad model spends less time in the low-luminosity phase compared to the Fidu model. Yet we caution that BH accretion is slightly suppressed when the radiation is stronger as discussed in Section~\ref{subsec:bh}. The higher AGN luminosity is caused directly by the higher radiative efficiency. From the ColdRadLow model to the ColdRadHigh model, the AGN has decreasing cumulative time above a given Eddington ratio owing to the declining BH accretion rate which overwhelms the effect of increasing radiative efficiency.  The bottom left panel exhibits the duty cycles of the BHmass model. The BHmass model shows more cumulative time of the AGN in the high-luminosity phase due to the lower value of Eddington luminosity.
The specific numbers of the AGN duty cycles above $0.02~L_\mathrm{Edd}$ are listed in the right column of Table~\ref{tab:effects}. It is obvious to find that the duty cycles of the cold-mode models have stronger variations than those of the hot-mode models.

The right panel of Figure~\ref{fig:dc} shows the fraction of total energy emitted above a given Eddington ratio.  Compared to the left panel, models reveal minor differences at low Eddington ratios, while the differences are amplified at high Eddington ratios. For the Fidu model, the AGN emits roughly $25\%$ of its total energy in the cold mode. This number is larger than the $6\%$ in the fullFB model of \citetalias{Yuan:2018ec}, which is caused by the correction of the bug that over-produces the energy flux of the wind in the cold mode. However, it is still not consistent with ``the Soltan argument'', which claims that AGNs emit most of their energy during the high-luminosity phase \citep{1982MNRAS.200..115S,2002MNRAS.335..965Y,Marconi2004}. As already discussed in \citetalias{Yuan:2018ec}, there are two main reasons for this discrepancy. The most important reason is that our simulations begin from 2~Gyr with a massive, mature elliptical galaxy. The central BH mass is already $10^9~M_\odot$ with minor growth in its subsequent evolution via accreting mass supplied by the stellar wind. In other words, the essential parts of the BH mass growth and the energy release occur before our simulations begin. The other reason is that we only focus on an isolated galaxy without considering the gaseous halo and cosmological inflow. The fraction of the energy emitted is expected to increase if we include the gas supply from the cosmological inflow and the gaseous halo.

Among all models, the fraction of cumulative energy emitted in the cold mode (the Eddington ratio $>10^{-2}~L_\mathrm{bol}/L_\mathrm{Edd}$) is the largest for the ColdWindFluxLow model, which is $\sim70\%$ of the total energy, while for the ColdWindFluxHigh model, almost zero energy is emitted in the cold mode. This suggests that the properties of wind in the cold mode are crucial to the high-luminosity phase of the galaxy. For the models studying the wind of the hot mode, it is interesting to find that although the time spent in the cold mode is roughly the same for the HotWindFluxHigh and HotWindFluxLow models, the AGN of the HotWindFluxHigh model emits a larger fraction of energy in the cold mode than that of the HotWindFluxLow model. This is because, on one hand, we find that the totally emitted energy in the cold mode in the two models are roughly the same; on the other hand, the typical AGN luminosity of the HotWindFluxHigh model in the hot mode is lower than that of the  HotWindFluxLow model thus the emitted energy in the hot mode of the HotWindFluxHigh model is lower.  This is also the reason for the larger fraction of the energy in the cold mode for the HotWindVel model.

In the center right panel, the AGN in the HotRad model emits a smaller fraction of energy in the cold mode than that of the Fidu model because of its higher total energy. The ColdRadLow model has a larger fraction of energy emitted in the cold mode due to its longer cumulative time spent in the cold mode. In particular, almost half of its energy is emitted in the cold mode for the ColdRadLow model, suggesting the non-negligible role of radiation. The ColdRadHigh model also has a larger fraction of energy emitted in the cold mode, although its AGN spends less time in the cold mode. This is caused by the higher radiative efficiency for the ColdRadHigh model. The bottom right panel presents the results for the BHmass model. The AGN of the BHmass model emits nearly half of the total energy in the cold mode.

\section{Summary and discussion}
\label{sec:discussions}

In \citetalias{Yuan:2018ec}, we have proposed a sub-grid model of AGN feedback and used it to study the AGN feedback in an elliptical galaxy. In that work, all the model parameters are set to be at their typical values. But some uncertainties still exist although black hole accretion is a relatively mature field compared to AGN feedback. In this paper, we perform  simulations to study the effects of parameters in the AGN sub-grid model of \citetalias{Yuan:2018ec}. Such a study is also useful for us to understand the role each model component plays in the feedback.  The fiducial model is the updated version of the fiducial model in \citetalias{Yuan:2018ec}. Based on this  model, we vary one parameter while keeping other parameters unchanged to study its effect. The models are listed in Table~\ref{tab:param}. By comparing models to the fiducial model, we are able to find the effect of this corresponding parameter on the AGN and galaxy evolution.

AGN Wind suppresses the BH accretion overall. Particularly, the wind velocity in the hot mode is the most important in controling the typical accretion rate and luminosity of the AGN. For example, a hot wind with three times higher velocity results in the typical AGN luminosity two orders of magnitude lower. When the accretion is in the cold mode, a powerful wind  stifles the accretion dramatically and renders the accretion back to the hot mode. This feedback-regulated BH emits at the typical luminosity between $10^{-6}~L_\mathrm{Edd}$ and $10^{-4}~L_\mathrm{Edd}$ for most of the time, which is consistent with the observations on the nearby, early-type galaxies \citep{2009ApJ...699..626H}. In the case of a stronger AGN wind, the mass growth of the BH decreases, and the AGN spends  a larger fraction of time in the low-luminosity phase. Star formation, however, takes place in a more complicated way. A direct correlation between the AGN wind power and the total star formation integrated over cosmological time and the whole galaxy is not found, which is because the range of wind  parameters explored here is not large enough.

AGN radiation also suppresses the BH accretion, although not as violently as the AGN wind. But it is premature to conclude that AGN wind plays a more important role than radiation in modulating BH accretion, because we have not considered dust in our simulations. Despite the negative effect on BH accretion, compared to the Fidu model, when the radiation of the hot mode becomes stronger, the AGN duty cycle shows a smaller fraction of time spent in the low-luminosity phase. This is caused by the larger radiative efficiency adopted. Similar to AGN wind, radiation of the hot mode also plays a role in reducing star formation at small radii while enhancing star formation at large radii. For radiation of the cold mode, however, stronger radiation suppresses the star formation at both small and large radii compared to the Fidu model.

Finally, we perform an extra model to investigate the effect of a lower initial BH mass, which is possibly suggested by the observations \citep{2014ApJ...789...17H,2015ApJ...813...82R,2016MNRAS.460.3119S,2019MNRAS.485.1278S}. A direct consequence is the stronger oscillations of the BH accretion rate and AGN luminosity. The AGN spends more time and emits more energy in the cold mode, and correspondingly the percentage of the BH mass growth increases. The star formation is also enhanced. 

In summary, however, given all the parameters we explore, the variations of the mass growth of BHs and the total star formation are within an order of magnitude, suggesting that our results are relatively insensitive to the parameters of the AGN sub-grid model.

In the initial state of our simulations, the galaxy already contains a very massive black hole and the star formation is already very low. We have shown in this paper (and our previous series of works, e.g., Yuan et al. 2018) that AGN feedback can keep the galaxy quenched. However, the gaseous halo and the cosmological inflow have so far been neglected. When these components are taken into account, it will be important to assess whether AGN feedback can still keep the galaxy quenched. This issue will be discussed in detail in our subsequent paper (Zhu et al. 2020, in preparation). Another issue is the effect of AGN feedback in the high redshift Universe. In that case, the black hole is small and the gas is abundant, so both the activity of the central AGN and star formation are much stronger. It is then interesting to ask what is the effect of AGN feedback; specifically, whether the AGN feedback can quench the galaxies. This question will be investigated in our future works.

\section*{Acknowledgements}
We thank our referee, Dr. Tiago Costa, for his constructive suggestions and comments. This work is supported in part by the National Key Research and Development Program of China (Grant No. 2016YFA0400704), the Natural Science Foundation of China (grants 11633006), and the Key Research Program of Frontier Sciences of CAS (No. QYZDJSSW-SYS008). This work made use of the High Performance Computing Resource in the Core Facility for Advanced Research Computing at Shanghai Astronomical Observatory.

\section*{Data Availability}
The data underlying this article will be shared on reasonable request to the corresponding author.




\bibliographystyle{mnras}
\bibliography{Parameter_survey} 

\begin{thebibliography}{}
\makeatletter
\relax
\def\mn@urlcharsother{\let\do\@makeother \do\$\do\&\do\#\do\^\do\_\do\%\do\~}
\def\mn@doi{\begingroup\mn@urlcharsother \@ifnextchar [ {\mn@doi@}
  {\mn@doi@[]}}
\def\mn@doi@[#1]#2{\def\@tempa{#1}\ifx\@tempa\@empty \href
  {http://dx.doi.org/#2} {doi:#2}\else \href {http://dx.doi.org/#2} {#1}\fi
  \endgroup}
\def\mn@eprint#1#2{\mn@eprint@#1:#2::\@nil}
\def\mn@eprint@arXiv#1{\href {http://arxiv.org/abs/#1} {{\tt arXiv:#1}}}
\def\mn@eprint@dblp#1{\href {http://dblp.uni-trier.de/rec/bibtex/#1.xml}
  {dblp:#1}}
\def\mn@eprint@#1:#2:#3:#4\@nil{\def\@tempa {#1}\def\@tempb {#2}\def\@tempc
  {#3}\ifx \@tempc \@empty \let \@tempc \@tempb \let \@tempb \@tempa \fi \ifx
  \@tempb \@empty \def\@tempb {arXiv}\fi \@ifundefined
  {mn@eprint@\@tempb}{\@tempb:\@tempc}{\expandafter \expandafter \csname
  mn@eprint@\@tempb\endcsname \expandafter{\@tempc}}}

\bibitem[\protect\citeauthoryear{{Arav}, {Korista}, {de Kool}, {Junkkarinen}
  \& {Begelman}}{{Arav} et~al.}{1999}]{1999ApJ...516...27A}
{Arav} N.,  {Korista} K.~T.,  {de Kool} M.,  {Junkkarinen} V.~T.,   {Begelman}
  M.~C.,  1999, \mn@doi [\apj] {10.1086/307073}, \href
  {https://ui.adsabs.harvard.edu/abs/1999ApJ...516...27A} {516, 27}

\bibitem[\protect\citeauthoryear{{Arav}, {Borguet}, {Chamberlain}, {Edmonds}
  \& {Danforth}}{{Arav} et~al.}{2013}]{2013MNRAS.436.3286A}
{Arav} N.,  {Borguet} B.,  {Chamberlain} C.,  {Edmonds} D.,   {Danforth} C.,
  2013, \mn@doi [\mnras] {10.1093/mnras/stt1812}, \href
  {https://ui.adsabs.harvard.edu/abs/2013MNRAS.436.3286A} {436, 3286}

\bibitem[\protect\citeauthoryear{{Bigiel}, {Leroy}, {Walter}, {Brinks}, {de
  Blok}, {Madore}  \& {Thornley}}{{Bigiel} et~al.}{2008}]{2008AJ....136.2846B}
{Bigiel} F.,  {Leroy} A.,  {Walter} F.,  {Brinks} E.,  {de Blok} W.~J.~G.,
  {Madore} B.,   {Thornley} M.~D.,  2008, \mn@doi [\aj]
  {10.1088/0004-6256/136/6/2846}, \href
  {https://ui.adsabs.harvard.edu/abs/2008AJ....136.2846B} {136, 2846}

\bibitem[\protect\citeauthoryear{Bower, Benson, Malbon, Helly, Frenk, Baugh,
  Cole  \& Lacey}{Bower et~al.}{2006}]{Bower2006}
Bower R.~G.,  Benson A.~J.,  Malbon R.,  Helly J.~C.,  Frenk C.~S.,  Baugh
  C.~M.,  Cole S.,   Lacey C.~G.,  2006, \mn@doi [\mnras]
  {10.1111/j.1365-2966.2006.10519.x}, 370, 645

\bibitem[\protect\citeauthoryear{Cappellari et~al.,}{Cappellari
  et~al.}{2013}]{Cappellari2013}
Cappellari M.,  et~al., 2013, \mn@doi [\mnras] {10.1093/mnras/stt644}, 432,
  1862

\bibitem[\protect\citeauthoryear{{Choi}, {Ostriker}, {Naab}  \&
  {Johansson}}{{Choi} et~al.}{2012}]{2012ApJ...754..125C}
{Choi} E.,  {Ostriker} J.~P.,  {Naab} T.,   {Johansson} P.~H.,  2012, \mn@doi
  [\apj] {10.1088/0004-637X/754/2/125}, \href
  {https://ui.adsabs.harvard.edu/abs/2012ApJ...754..125C} {754, 125}

\bibitem[\protect\citeauthoryear{{Choi}, {Ostriker}, {Naab}, {Oser}  \&
  {Moster}}{{Choi} et~al.}{2015}]{2015MNRAS.449.4105C}
{Choi} E.,  {Ostriker} J.~P.,  {Naab} T.,  {Oser} L.,   {Moster} B.~P.,  2015,
  \mn@doi [\mnras] {10.1093/mnras/stv575}, \href
  {https://ui.adsabs.harvard.edu/abs/2015MNRAS.449.4105C} {449, 4105}

\bibitem[\protect\citeauthoryear{Ciotti \& Ostriker}{Ciotti \&
  Ostriker}{1997}]{Ciotti1997}
Ciotti L.,  Ostriker J.~P.,  1997, \mn@doi [\apj] {10.1086/310902}, 487, L105

\bibitem[\protect\citeauthoryear{Ciotti \& Ostriker}{Ciotti \&
  Ostriker}{2001}]{Ciotti2001}
Ciotti L.,  Ostriker J.~P.,  2001, \mn@doi [\apj] {10.1086/320053}, 551, 131

\bibitem[\protect\citeauthoryear{Ciotti \& Ostriker}{Ciotti \&
  Ostriker}{2007}]{Ciotti2007}
Ciotti L.,  Ostriker J.~P.,  2007, \mn@doi [\apj] {10.1086/519833}, 665, 1038

\bibitem[\protect\citeauthoryear{{Ciotti} \& {Ostriker}}{{Ciotti} \&
  {Ostriker}}{2012}]{2012ASSL..378...83C}
{Ciotti} L.,  {Ostriker} J.~P.,  2012, in {Kim} D.-W.,  {Pellegrini} S.,  eds,
  Astrophysics and Space Science Library Vol. 378, Astrophysics and Space
  Science Library. p.~83 (\mn@eprint {arXiv} {1104.2238}),
  \mn@doi{10.1007/978-1-4614-0580-1_4}

\bibitem[\protect\citeauthoryear{Ciotti, Ostriker  \& Proga}{Ciotti
  et~al.}{2009}]{Ciotti2009}
Ciotti L.,  Ostriker J.~P.,   Proga D.,  2009, \mn@doi [\apj]
  {10.1088/0004-637X/699/1/89}, 699, 89

\bibitem[\protect\citeauthoryear{Ciotti, Ostriker  \& Proga}{Ciotti
  et~al.}{2010}]{Ciotti2010}
Ciotti L.,  Ostriker J.~P.,   Proga D.,  2010, \mn@doi [\apj]
  {10.1088/0004-637X/717/2/708}, 717, 708

\bibitem[\protect\citeauthoryear{{Ciotti}, {Pellegrini}, {Negri}  \&
  {Ostriker}}{{Ciotti} et~al.}{2017}]{2017ApJ...835...15C}
{Ciotti} L.,  {Pellegrini} S.,  {Negri} A.,   {Ostriker} J.~P.,  2017, \mn@doi
  [\apj] {10.3847/1538-4357/835/1/15}, \href
  {https://ui.adsabs.harvard.edu/abs/2017ApJ...835...15C} {835, 15}

\bibitem[\protect\citeauthoryear{{Costa}, {Rosdahl}, {Sijacki}  \&
  {Haehnelt}}{{Costa} et~al.}{2018}]{2018MNRAS.473.4197C}
{Costa} T.,  {Rosdahl} J.,  {Sijacki} D.,   {Haehnelt} M.~G.,  2018, \mn@doi
  [\mnras] {10.1093/mnras/stx2598}, \href
  {https://ui.adsabs.harvard.edu/abs/2018MNRAS.473.4197C} {473, 4197}

\bibitem[\protect\citeauthoryear{Croton et~al.,}{Croton
  et~al.}{2006}]{Croton2006}
Croton D.~J.,  et~al., 2006, \mn@doi [\mnras]
  {10.1111/j.1365-2966.2005.09675.x}, 365, 11

\bibitem[\protect\citeauthoryear{{Cui} \& {Yuan}}{{Cui} \&
  {Yuan}}{2020}]{2020ApJ...890...81C}
{Cui} C.,  {Yuan} F.,  2020, \mn@doi [\apj] {10.3847/1538-4357/ab6e6f}, \href
  {https://ui.adsabs.harvard.edu/abs/2020ApJ...890...81C} {890, 81}

\bibitem[\protect\citeauthoryear{{Cui}, {Yuan}  \& {Li}}{{Cui}
  et~al.}{2020}]{2020ApJ...890...80C}
{Cui} C.,  {Yuan} F.,   {Li} B.,  2020, \mn@doi [\apj]
  {10.3847/1538-4357/ab6e6e}, \href
  {https://ui.adsabs.harvard.edu/abs/2020ApJ...890...80C} {890, 80}

\bibitem[\protect\citeauthoryear{{Di Matteo}, Springel  \& Ilernquist}{{Di
  Matteo} et~al.}{2005}]{DiMatteo2005}
{Di Matteo} T.,  Springel V.,   Ilernquist L.,  2005, \mn@doi [\nat]
  {10.1038/nature03335}, 433, 604

\bibitem[\protect\citeauthoryear{{Dubois} et~al.,}{{Dubois}
  et~al.}{2014}]{2014MNRAS.444.1453D}
{Dubois} Y.,  et~al., 2014, \mn@doi [\mnras] {10.1093/mnras/stu1227}, \href
  {https://ui.adsabs.harvard.edu/abs/2014MNRAS.444.1453D} {444, 1453}

\bibitem[\protect\citeauthoryear{{Ferrarese} \& {Merritt}}{{Ferrarese} \&
  {Merritt}}{2000}]{2000ApJ...539L...9F}
{Ferrarese} L.,  {Merritt} D.,  2000, \mn@doi [\apjl] {10.1086/312838}, \href
  {https://ui.adsabs.harvard.edu/abs/2000ApJ...539L...9F} {539, L9}

\bibitem[\protect\citeauthoryear{{Frank}, {King}  \& {Raine}}{{Frank}
  et~al.}{2002}]{2002apa..book.....F}
{Frank} J.,  {King} A.,   {Raine} D.~J.,  2002, {Accretion Power in
  Astrophysics: Third Edition}.
Cambridge University Press

\bibitem[\protect\citeauthoryear{Gan, Yuan, Ostriker, Ciotti  \& Novak}{Gan
  et~al.}{2014}]{Gan2014}
Gan Z.,  Yuan F.,  Ostriker J.~P.,  Ciotti L.,   Novak G.~S.,  2014, \mn@doi
  [\apj] {10.1088/0004-637X/789/2/150}, 789, 150

\bibitem[\protect\citeauthoryear{Gan, Ciotti, Ostriker  \& Yuan}{Gan
  et~al.}{2019}]{Gan2019}
Gan Z.,  Ciotti L.,  Ostriker J.~P.,   Yuan F.,  2019, \mn@doi [\apj]
  {10.3847/1538-4357/ab0206}, 872, 167

\bibitem[\protect\citeauthoryear{Gebhardt et~al.,}{Gebhardt
  et~al.}{2000}]{Gebhardt2000}
Gebhardt K.,  et~al., 2000, \mn@doi [\apj] {10.1086/318174}, 543, L5

\bibitem[\protect\citeauthoryear{Gofford, Reeves, Tombesi, Braito, Turner,
  Miller  \& Cappi}{Gofford et~al.}{2013}]{Gofford2013}
Gofford J.,  Reeves J.~N.,  Tombesi F.,  Braito V.,  Turner T.~J.,  Miller L.,
   Cappi M.,  2013, \mn@doi [\mnras] {10.1093/mnras/sts481}, 430, 60

\bibitem[\protect\citeauthoryear{Gofford, Reeves, McLaughlin, Braito, Turner,
  Tombesi  \& Cappi}{Gofford et~al.}{2015}]{2015MNRAS.451.4169G}
Gofford J.,  Reeves J.~N.,  McLaughlin D.~E.,  Braito V.,  Turner T.~J.,
  Tombesi F.,   Cappi M.,  2015, \mn@doi [\mnras] {10.1093/mnras/stv1207}, 451,
  4169

\bibitem[\protect\citeauthoryear{{Graham} et~al.,}{{Graham}
  et~al.}{2018}]{2018MNRAS.477.4711G}
{Graham} M.~T.,  et~al., 2018, \mn@doi [\mnras] {10.1093/mnras/sty504}, \href
  {https://ui.adsabs.harvard.edu/abs/2018MNRAS.477.4711G} {477, 4711}

\bibitem[\protect\citeauthoryear{{G{\"u}ltekin} et~al.,}{{G{\"u}ltekin}
  et~al.}{2009}]{2009ApJ...698..198G}
{G{\"u}ltekin} K.,  et~al., 2009, \mn@doi [\apj] {10.1088/0004-637X/698/1/198},
  \href {https://ui.adsabs.harvard.edu/abs/2009ApJ...698..198G} {698, 198}

\bibitem[\protect\citeauthoryear{{H{\"a}ring} \& {Rix}}{{H{\"a}ring} \&
  {Rix}}{2004}]{2004ApJ...604L..89H}
{H{\"a}ring} N.,  {Rix} H.-W.,  2004, \mn@doi [\apjl] {10.1086/383567}, \href
  {https://ui.adsabs.harvard.edu/abs/2004ApJ...604L..89H} {604, L89}

\bibitem[\protect\citeauthoryear{{Harrison}, {Costa}, {Tadhunter},
  {Fl{\"u}tsch}, {Kakkad}, {Perna}  \& {Vietri}}{{Harrison}
  et~al.}{2018}]{2018NatAs...2..198H}
{Harrison} C.~M.,  {Costa} T.,  {Tadhunter} C.~N.,  {Fl{\"u}tsch} A.,  {Kakkad}
  D.,  {Perna} M.,   {Vietri} G.,  2018, \mn@doi [Nature Astronomy]
  {10.1038/s41550-018-0403-6}, \href
  {https://ui.adsabs.harvard.edu/abs/2018NatAs...2..198H} {2, 198}

\bibitem[\protect\citeauthoryear{Hayes, Norman, Fiedler, Bordner, Li, Clark, ud
  Doula  \& Mac~Low}{Hayes et~al.}{2006}]{2006ApJS..165..188H}
Hayes J.~C.,  Norman M.~L.,  Fiedler R.~A.,  Bordner J.~O.,  Li P.~S.,  Clark
  S.~E.,  ud Doula A.,   Mac~Low M.~M.,  2006, \mn@doi [\apjs]
  {10.1086/504594}, 165, 188

\bibitem[\protect\citeauthoryear{Ho}{Ho}{2009}]{2009ApJ...699..626H}
Ho L.~C.,  2009, \mn@doi [\apj] {10.1088/0004-637X/699/1/626}, 699, 626

\bibitem[\protect\citeauthoryear{{Ho} \& {Kim}}{{Ho} \&
  {Kim}}{2014}]{2014ApJ...789...17H}
{Ho} L.~C.,  {Kim} M.,  2014, \mn@doi [\apj] {10.1088/0004-637X/789/1/17},
  \href {https://ui.adsabs.harvard.edu/abs/2014ApJ...789...17H} {789, 17}

\bibitem[\protect\citeauthoryear{{Jaffe}}{{Jaffe}}{1983}]{1983MNRAS.202..995J}
{Jaffe} W.,  1983, \mn@doi [\mnras] {10.1093/mnras/202.4.995}, \href
  {https://ui.adsabs.harvard.edu/abs/1983MNRAS.202..995J} {202, 995}

\bibitem[\protect\citeauthoryear{{Kato}, {Fukue}  \& {Mineshige}}{{Kato}
  et~al.}{2008}]{2008bhad.book.....K}
{Kato} S.,  {Fukue} J.,   {Mineshige} S.,  2008, {Black-Hole Accretion Disks
  --- Towards a New Paradigm ---}.
Kyoto University Press

\bibitem[\protect\citeauthoryear{{Kennicutt}}{{Kennicutt}}{1989}]{1989ApJ...344..685K}
{Kennicutt} Robert~C. J.,  1989, \mn@doi [\apj] {10.1086/167834}, \href
  {https://ui.adsabs.harvard.edu/abs/1989ApJ...344..685K} {344, 685}

\bibitem[\protect\citeauthoryear{{Kennicutt}}{{Kennicutt}}{1998}]{1998ApJ...498..541K}
{Kennicutt} Robert~C. J.,  1998, \mn@doi [\apj] {10.1086/305588}, \href
  {https://ui.adsabs.harvard.edu/abs/1998ApJ...498..541K} {498, 541}

\bibitem[\protect\citeauthoryear{{Khandai}, {Di Matteo}, {Croft}, {Wilkins},
  {Feng}, {Tucker}, {DeGraf}  \& {Liu}}{{Khandai}
  et~al.}{2015}]{2015MNRAS.450.1349K}
{Khandai} N.,  {Di Matteo} T.,  {Croft} R.,  {Wilkins} S.,  {Feng} Y.,
  {Tucker} E.,  {DeGraf} C.,   {Liu} M.-S.,  2015, \mn@doi [\mnras]
  {10.1093/mnras/stv627}, \href
  {https://ui.adsabs.harvard.edu/abs/2015MNRAS.450.1349K} {450, 1349}

\bibitem[\protect\citeauthoryear{{King} \& {Pounds}}{{King} \&
  {Pounds}}{2003}]{2003MNRAS.345..657K}
{King} A.~R.,  {Pounds} K.~A.,  2003, \mn@doi [\mnras]
  {10.1046/j.1365-8711.2003.06980.x}, \href
  {https://ui.adsabs.harvard.edu/abs/2003MNRAS.345..657K} {345, 657}

\bibitem[\protect\citeauthoryear{{King} \& {Pounds}}{{King} \&
  {Pounds}}{2015}]{2015ARA&A..53..115K}
{King} A.,  {Pounds} K.,  2015, \mn@doi [\araa]
  {10.1146/annurev-astro-082214-122316}, \href
  {https://ui.adsabs.harvard.edu/abs/2015ARA&A..53..115K} {53, 115}

\bibitem[\protect\citeauthoryear{Kormendy \& Ho}{Kormendy \&
  Ho}{2013}]{Kormendy_Coevolution_2013}
Kormendy J.,  Ho L.~C.,  2013, \mn@doi [\araa]
  {10.1146/annurev-astro-082708-101811}, 51, 511

\bibitem[\protect\citeauthoryear{{Krumholz} \& {Tan}}{{Krumholz} \&
  {Tan}}{2007}]{2007ApJ...654..304K}
{Krumholz} M.~R.,  {Tan} J.~C.,  2007, \mn@doi [\apj] {10.1086/509101}, \href
  {https://ui.adsabs.harvard.edu/abs/2007ApJ...654..304K} {654, 304}

\bibitem[\protect\citeauthoryear{{Krumholz}, {McKee}  \& {Bland
  -Hawthorn}}{{Krumholz} et~al.}{2019}]{2019ARA&A..57..227K}
{Krumholz} M.~R.,  {McKee} C.~F.,   {Bland -Hawthorn} J.,  2019, \mn@doi
  [\araa] {10.1146/annurev-astro-091918-104430}, \href
  {https://ui.adsabs.harvard.edu/abs/2019ARA&A..57..227K} {57, 227}

\bibitem[\protect\citeauthoryear{Magorrian et~al.,}{Magorrian
  et~al.}{1998}]{Magorrian1998}
Magorrian J.,  et~al., 1998, \mn@doi [\apj] {10.1086/300353}, 115, 2285

\bibitem[\protect\citeauthoryear{{Marconi} \& {Hunt}}{{Marconi} \&
  {Hunt}}{2003}]{2003ApJ...589L..21M}
{Marconi} A.,  {Hunt} L.~K.,  2003, \mn@doi [\apjl] {10.1086/375804}, \href
  {https://ui.adsabs.harvard.edu/abs/2003ApJ...589L..21M} {589, L21}

\bibitem[\protect\citeauthoryear{Marconi, Risaliti, Gilli, Hunt, Maiolino  \&
  Salvati}{Marconi et~al.}{2004}]{Marconi2004}
Marconi A.,  Risaliti G.,  Gilli R.,  Hunt L.~K.,  Maiolino R.,   Salvati M.,
  2004, \mn@doi [\mnras] {10.1111/j.1365-2966.2004.07765.x}, 351, 169

\bibitem[\protect\citeauthoryear{{Martin} \& {Kennicutt}}{{Martin} \&
  {Kennicutt}}{2001}]{2001ApJ...555..301M}
{Martin} C.~L.,  {Kennicutt} Robert~C. J.,  2001, \mn@doi [\apj]
  {10.1086/321452}, \href
  {https://ui.adsabs.harvard.edu/abs/2001ApJ...555..301M} {555, 301}

\bibitem[\protect\citeauthoryear{{Monaco}, {Fontanot}  \& {Taffoni}}{{Monaco}
  et~al.}{2007}]{2007MNRAS.375.1189M}
{Monaco} P.,  {Fontanot} F.,   {Taffoni} G.,  2007, \mn@doi [\mnras]
  {10.1111/j.1365-2966.2006.11253.x}, \href
  {https://ui.adsabs.harvard.edu/abs/2007MNRAS.375.1189M} {375, 1189}

\bibitem[\protect\citeauthoryear{Negri \& Volonteri}{Negri \&
  Volonteri}{2017}]{Negri2017}
Negri A.,  Volonteri M.,  2017, \mn@doi [\mnras] {10.1093/mnras/stx362}, 467,
  3475

\bibitem[\protect\citeauthoryear{Novak, Ostriker  \& Ciotti}{Novak
  et~al.}{2011}]{Novak2011}
Novak G.~S.,  Ostriker J.~P.,   Ciotti L.,  2011, \mn@doi [\apj]
  {10.1088/0004-637X/737/1/26}, 737, 14

\bibitem[\protect\citeauthoryear{{Pillepich} et~al.,}{{Pillepich}
  et~al.}{2018}]{2018MNRAS.473.4077P}
{Pillepich} A.,  et~al., 2018, \mn@doi [\mnras] {10.1093/mnras/stx2656}, \href
  {https://ui.adsabs.harvard.edu/abs/2018MNRAS.473.4077P} {473, 4077}

\bibitem[\protect\citeauthoryear{{Reines} \& {Volonteri}}{{Reines} \&
  {Volonteri}}{2015}]{2015ApJ...813...82R}
{Reines} A.~E.,  {Volonteri} M.,  2015, \mn@doi [\apj]
  {10.1088/0004-637X/813/2/82}, \href
  {https://ui.adsabs.harvard.edu/abs/2015ApJ...813...82R} {813, 82}

\bibitem[\protect\citeauthoryear{{Sazonov}, {Ostriker}  \& {Sunyaev}}{{Sazonov}
  et~al.}{2004}]{2004MNRAS.347..144S}
{Sazonov} S.~Y.,  {Ostriker} J.~P.,   {Sunyaev} R.~A.,  2004, \mn@doi [\mnras]
  {10.1111/j.1365-2966.2004.07184.x}, \href
  {https://ui.adsabs.harvard.edu/abs/2004MNRAS.347..144S} {347, 144}

\bibitem[\protect\citeauthoryear{{Sazonov}, {Ostriker}, {Ciotti}  \&
  {Sunyaev}}{{Sazonov} et~al.}{2005}]{2005MNRAS.358..168S}
{Sazonov} S.~Y.,  {Ostriker} J.~P.,  {Ciotti} L.,   {Sunyaev} R.~A.,  2005,
  \mn@doi [\mnras] {10.1111/j.1365-2966.2005.08763.x}, \href
  {https://ui.adsabs.harvard.edu/abs/2005MNRAS.358..168S} {358, 168}

\bibitem[\protect\citeauthoryear{{Schaye} et~al.,}{{Schaye}
  et~al.}{2015}]{2015MNRAS.446..521S}
{Schaye} J.,  et~al., 2015, \mn@doi [\mnras] {10.1093/mnras/stu2058}, \href
  {https://ui.adsabs.harvard.edu/abs/2015MNRAS.446..521S} {446, 521}

\bibitem[\protect\citeauthoryear{{Shankar} et~al.,}{{Shankar}
  et~al.}{2016}]{2016MNRAS.460.3119S}
{Shankar} F.,  et~al., 2016, \mn@doi [\mnras] {10.1093/mnras/stw678}, \href
  {https://ui.adsabs.harvard.edu/abs/2016MNRAS.460.3119S} {460, 3119}

\bibitem[\protect\citeauthoryear{{Shankar} et~al.,}{{Shankar}
  et~al.}{2019}]{2019MNRAS.485.1278S}
{Shankar} F.,  et~al., 2019, \mn@doi [\mnras] {10.1093/mnras/stz376}, \href
  {https://ui.adsabs.harvard.edu/abs/2019MNRAS.485.1278S} {485, 1278}

\bibitem[\protect\citeauthoryear{Shin, Ostriker  \& Ciotti}{Shin
  et~al.}{2010}]{Shin2010}
Shin M.~S.,  Ostriker J.~P.,   Ciotti L.,  2010, \mn@doi [\apj]
  {10.1088/0004-637X/711/1/268}, 711, 268

\bibitem[\protect\citeauthoryear{{Soltan}}{{Soltan}}{1982}]{1982MNRAS.200..115S}
{Soltan} A.,  1982, \mn@doi [\mnras] {10.1093/mnras/200.1.115}, \href
  {https://ui.adsabs.harvard.edu/abs/1982MNRAS.200..115S} {200, 115}

\bibitem[\protect\citeauthoryear{Springel, {Di Matteo}  \& Hernquist}{Springel
  et~al.}{2005}]{Springel2005}
Springel V.,  {Di Matteo} T.,   Hernquist L.,  2005, \mn@doi [\mnras]
  {10.1111/j.1365-2966.2005.09238.x}, 361, 776

\bibitem[\protect\citeauthoryear{{Tombesi}, {Cappi}, {Reeves}, {Palumbo},
  {Yaqoob}, {Braito}  \& {Dadina}}{{Tombesi}
  et~al.}{2010}]{2010A&A...521A..57T}
{Tombesi} F.,  {Cappi} M.,  {Reeves} J.~N.,  {Palumbo} G.~G.~C.,  {Yaqoob} T.,
  {Braito} V.,   {Dadina} M.,  2010, \mn@doi [\aap]
  {10.1051/0004-6361/200913440}, \href
  {https://ui.adsabs.harvard.edu/abs/2010A&A...521A..57T} {521, A57}

\bibitem[\protect\citeauthoryear{{Tombesi}, {Cappi}, {Reeves}  \&
  {Braito}}{{Tombesi} et~al.}{2012}]{2012MNRAS.422L...1T}
{Tombesi} F.,  {Cappi} M.,  {Reeves} J.~N.,   {Braito} V.,  2012, \mn@doi
  [\mnras] {10.1111/j.1745-3933.2012.01221.x}, \href
  {https://ui.adsabs.harvard.edu/abs/2012MNRAS.422L...1T} {422, L1}

\bibitem[\protect\citeauthoryear{{Tremaine} et~al.,}{{Tremaine}
  et~al.}{2002}]{2002ApJ...574..740T}
{Tremaine} S.,  et~al., 2002, \mn@doi [\apj] {10.1086/341002}, \href
  {https://ui.adsabs.harvard.edu/abs/2002ApJ...574..740T} {574, 740}

\bibitem[\protect\citeauthoryear{{Vogelsberger} et~al.,}{{Vogelsberger}
  et~al.}{2014}]{2014MNRAS.444.1518V}
{Vogelsberger} M.,  et~al., 2014, \mn@doi [\mnras] {10.1093/mnras/stu1536},
  \href {https://ui.adsabs.harvard.edu/abs/2014MNRAS.444.1518V} {444, 1518}

\bibitem[\protect\citeauthoryear{Xie \& Yuan}{Xie \& Yuan}{2012}]{Xie:2012dv}
Xie F.-G.,  Yuan F.,  2012, \mn@doi [\mnras]
  {10.1111/j.1365-2966.2012.22030.x}, 427, 1580

\bibitem[\protect\citeauthoryear{{Xie}, {Yuan}  \& {Ho}}{{Xie}
  et~al.}{2017}]{2017ApJ...844...42X}
{Xie} F.-G.,  {Yuan} F.,   {Ho} L.~C.,  2017, \mn@doi [\apj]
  {10.3847/1538-4357/aa7950}, \href
  {https://ui.adsabs.harvard.edu/abs/2017ApJ...844...42X} {844, 42}

\bibitem[\protect\citeauthoryear{Yoon, Yuan, Gan, Ostriker, Li  \& Ciotti}{Yoon
  et~al.}{2018}]{Yoon2018}
Yoon D.,  Yuan F.,  Gan Z.-M.,  Ostriker J.~P.,  Li Y.-P.,   Ciotti L.,  2018,
  \mn@doi [\apj] {10.3847/1538-4357/aad37e}, 864, 6

\bibitem[\protect\citeauthoryear{{Yoon}, {Yuan}, {Ostriker}, {Ciotti}  \&
  {Zhu}}{{Yoon} et~al.}{2019}]{2019ApJ...885...16Y}
{Yoon} D.,  {Yuan} F.,  {Ostriker} J.~P.,  {Ciotti} L.,   {Zhu} B.,  2019,
  \mn@doi [\apj] {10.3847/1538-4357/ab45e8}, \href
  {https://ui.adsabs.harvard.edu/abs/2019ApJ...885...16Y} {885, 16}

\bibitem[\protect\citeauthoryear{{Yu} \& {Tremaine}}{{Yu} \&
  {Tremaine}}{2002}]{2002MNRAS.335..965Y}
{Yu} Q.,  {Tremaine} S.,  2002, \mn@doi [\mnras]
  {10.1046/j.1365-8711.2002.05532.x}, \href
  {https://ui.adsabs.harvard.edu/abs/2002MNRAS.335..965Y} {335, 965}

\bibitem[\protect\citeauthoryear{Yuan \& Narayan}{Yuan \&
  Narayan}{2014}]{2014ARA&A..52..529Y}
Yuan F.,  Narayan R.,  2014, \mn@doi [\araa]
  {10.1146/annurev-astro-082812-141003}, 52, 529

\bibitem[\protect\citeauthoryear{{Yuan}, {Bu}  \& {Wu}}{{Yuan}
  et~al.}{2012}]{2012ApJ...761..130Y}
{Yuan} F.,  {Bu} D.,   {Wu} M.,  2012, \mn@doi [\apj]
  {10.1088/0004-637X/761/2/130}, \href
  {https://ui.adsabs.harvard.edu/abs/2012ApJ...761..130Y} {761, 130}

\bibitem[\protect\citeauthoryear{{Yuan}, {Gan}, {Narayan}, {Sadowski}, {Bu}  \&
  {Bai}}{{Yuan} et~al.}{2015}]{2015ApJ...804..101Y}
{Yuan} F.,  {Gan} Z.,  {Narayan} R.,  {Sadowski} A.,  {Bu} D.,   {Bai} X.-N.,
  2015, \mn@doi [\apj] {10.1088/0004-637X/804/2/101}, \href
  {https://ui.adsabs.harvard.edu/abs/2015ApJ...804..101Y} {804, 101}

\bibitem[\protect\citeauthoryear{Yuan, Yoon, Li, Gan, Ho  \& Guo}{Yuan
  et~al.}{2018}]{Yuan:2018ec}
Yuan F.,  Yoon D.,  Li Y.-P.,  Gan Z.-M.,  Ho L.~C.,   Guo F.,  2018, \mn@doi
  [\apj] {10.3847/1538-4357/aab8f8}, 857, 0

\bibitem[\protect\citeauthoryear{{Zeilig-Hess}, {Levinson}  \&
  {Nakar}}{{Zeilig-Hess} et~al.}{2019}]{2019MNRAS.482.4642Z}
{Zeilig-Hess} M.,  {Levinson} A.,   {Nakar} E.,  2019, \mn@doi [\mnras]
  {10.1093/mnras/sty3034}, \href
  {https://ui.adsabs.harvard.edu/abs/2019MNRAS.482.4642Z} {482, 4642}

\makeatother
\end{thebibliography}





%


\bsp	
\label{lastpage}
\end{document}